\documentclass[sigconf]{acmart}

\usepackage{booktabs}
\usepackage{enumitem}
\usepackage{multirow}
\usepackage{graphicx}
\usepackage{xspace}
\usepackage{fontawesome}
\usepackage{csquotes}
\usepackage{dsfont}
\usepackage{makecell}
\usepackage{framed, color}
\usepackage{soul}
\usepackage{subcaption}

\newcommand{\eg}{{e.g.}, }
\newcommand{\ie}{{i.e.}, }
\newcommand{\etal}{{et al.}~}
\newcommand{\vs}{{vs.} }

\newcommand{\one}[0] {{(i)}~}
\newcommand{\two}[0] {{(ii)}~}
\newcommand{\three}[0] {{(iii)}~}

\newcommand{\nap}[0]{{\sc NaviPath}}

\AtBeginDocument{%
  \providecommand\BibTeX{{%
    \normalfont B\kern-0.5em{\scshape i\kern-0.25em b}\kern-0.8em\TeX}}}

\copyrightyear{2023}
\acmYear{2023}
\setcopyright{rightsretained}
\acmConference[CHI '23]{Proceedings of the 2023 CHI Conference on
Human Factors in Computing Systems}{April 23--28, 2023}{Hamburg,
Germany}
\acmBooktitle{Proceedings of the 2023 CHI Conference on Human Factors
in Computing Systems (CHI '23), April 23--28, 2023, Hamburg,
Germany}\acmDOI{10.1145/3544548.3580694}
\acmISBN{978-1-4503-9421-5/23/04}

\author{Hongyan Gu}
\email{ghy@ucla.edu}
\orcid{0000-0001-8962-9152}
\affiliation{%
  \streetaddress{580 Portola Plaza, Room 1538}
  \institution{University of California, Los Angeles}
  \city{Los Angeles}
  \state{California}
  \postcode{90095}
  \country{USA}}

\author{Chunxu Yang}
\email{chunxuyang@ucla.edu}
\orcid{0000-0002-9684-7534}
\affiliation{%
  \institution{University of California, Los Angeles}
  \city{Los Angeles}
  \state{California}
  \postcode{90095}
  \country{USA}
}

\author{Mohammad Haeri}
\email{mhaeri@kumc.edu}
\orcid{0000-0001-6055-9779}
\affiliation{%
  \institution{University of Kansas Medical Center}
  \city{Kansas City}
  \state{Kansas}
  \postcode{66160}
  \country{USA}}

\author{Jing Wang}
\email{jingwang829@outlook.com}
\orcid{0000-0003-2751-9868}
\affiliation{%
  \institution{Beijing Tongren Hospital, Capital Medical University}
  \city{Beijing}
  \postcode{100730}
  \country{China}}

\author{Shirley Tang}
\email{sjwtang@ucla.edu}
\orcid{0000-0002-0984-5208}
\affiliation{%
  \institution{University of California, Los Angeles}
  \city{Los Angeles}
  \state{California}
  \postcode{90095}
  \country{USA}
}

\author{Wenzhong Yan}
\orcid{0000-0002-3711-0807}
\email{wzyan24@ucla.edu}
\affiliation{%
  \institution{University of California, Los Angeles}
  \city{Los Angeles}
  \state{California}
  \postcode{90095}
  \country{USA}
}

\author{Shujin He}
\email{heshujun@mail.jnmc.edu.cn}
\orcid{0000-0003-3577-6346}
\affiliation{%
  \institution{Beijing Tongren Hospital, Capital Medical University}
  \city{Beijing}
  \postcode{100730}
  \country{China}}

\author{Christopher Kazu Williams}
\email{ckwilliams@mednet.ucla.edu}
\orcid{0000-0003-3253-7630}
\affiliation{%
  \institution{UCLA David Geffen School of Medicine}
  \city{Los Angeles}
  \state{California}
  \country{USA}
  \postcode{90095}
}

\author{Shino Magaki}
\email{smagaki@mednet.ucla.edu}
\orcid{0000-0003-0433-5759}
\affiliation{%
  \institution{UCLA David Geffen School of Medicine}
  \city{Los Angeles}
  \state{California}
  \country{USA}
  \postcode{90095}
}

\author{Xiang `Anthony' Chen}
\email{xac@ucla.edu}
\orcid{0000-0002-8527-1744}
\affiliation{%
  \institution{University of California, Los Angeles}
  \city{Los Angeles}
  \state{California}
  \postcode{90095}
  \country{USA}}

\renewcommand{\shortauthors}{Gu, et al.}

\begin{document}

\title[NaviPath]{Augmenting Pathologists with NaviPath: Design and Evaluation of a Human-AI Collaborative Navigation System}


\begin{abstract}
Artificial Intelligence (AI) brings advancements to support pathologists in navigating high-resolution tumor images to search for pathology patterns of interest. However, existing AI-assisted tools have not realized this promised potential due to a lack of insight into pathology and HCI considerations for pathologists' navigation workflows in practice. We first conducted a formative study with six medical professionals in pathology to capture their navigation strategies. By incorporating our observations along with the pathologists' domain knowledge, we designed \nap~--- a human-AI collaborative navigation system. An evaluation study with 15 medical professionals in pathology indicated that: \one compared to the manual navigation, participants saw more than twice the number of pathological patterns in unit time with \nap, and \two participants achieved higher precision and recall against the AI and the manual navigation on average. Further qualitative analysis revealed that navigation was more consistent with \nap, which can improve the overall examination quality.
\end{abstract}



\sloppy
\maketitle


\section{Introduction}

One crucial step of cancer diagnoses is the pathologists' examinations of tumors through an optical microscope. With the recent development of digital pathology \cite{fdawsi2017, pantanowitz2011review}, tumor specimens can be scanned into high-resolution digital scans, allowing medical professionals to access, analyze, and share these scans with digital interfaces \cite{martel2017image, gutman2017digital, saltz2017containerized}. However, literature has suggested that it might take longer for pathologists to examine digital scans compared to when using microscopes \cite{treanor2007virtual, hanna2019whole}. The main culprit is the difficulty in navigation --- pathology scans usually have extremely high resolutions ($(\sim 10^6)^2$ pixels) compared to commercial off-the-shelf computer displays ($\sim 8.3 \times 10^6$ pixels for 4K UHD resolution). Therefore, pathologists are required to frequently manipulate (\ie zooming, panning) the viewport to gather necessary information for diagnoses \cite{ruddle2016design}. 

Research has long realized the difficulty in navigating high-resolution images and proposed various interface designs to assist users with general navigation tasks (\eg map exploration) \cite{bederson1996pad++, zellweger2003city, gustafson2008wedge, sarkar1992graphical, cockburn2009review}. However, we believe necessary adaptations should be considered to enable seamless integration into pathologists' workflows, because of three problems in human navigation of pathology scans: \one pathologists' navigation is usually substantially complicated because some pathology patterns (\eg mitosis in low-grade meningiomas \cite{louis20212021}) have a low prevalence rate (<100/scan) and have extremely small dimensions compared to pathology scans (ratio up to 1:2000) \cite{aubreville2020completely}; \two pathologists require specific domain knowledge and navigation strategies \cite{ruddle2016design, molin2015slide} to facilitate their examinations, which current navigation systems for general use rarely consider; \three although AI can be used to accelerate navigation, the lack of consideration towards integrating AI into pathologists' workflows might discourage them from using human-AI systems in practice, as suggested in previous studies \cite{yang2016investigating, gu2021xpath}. Fortunately, recent HCI-AI-Health works have demonstrated prototypes and designs to close the gap between medical professionals and AI, which has facilitated human-AI communication and was viable to improve doctors' works in various medical application domains, such as general medicine \cite{Yang2019, lee2021human, schaekermann2020ambiguity}, radiology \cite{calisto2022breastscreening, calisto2021introduction} and pathology \cite{cai2019human, lindvall2021rapid, gu2021xpath}. Motivated by the success of these advancements, this work continues to build integrable systems by taking doctors' domain knowledge into account, with a focus on supporting the navigation process in pathology.

To this end, we conducted a formative study with six medical professionals in pathology from two medical centers to enrich our understanding of their navigation processes. Specifically, we observed how they navigated pathology scans to search for mitoses\footnote{The mitosis is selected because \one the size of mitoses is small ($\sim 10\mu m$) compared to the size of pathology scans; \two the prevalence of mitoses is low ($<0.2/(1,600)^2$ pixels in specific carcinomas) \cite{aubreville2020completely}.}, a critical pathology pattern that relates to cancer malignancy and patient prognosis \cite{cree2021counting}. We summarized three observations that cross-validate the findings in previous research \cite{545307, glueck2013model, ruddle2016design, molin2015slide}:

\begin{figure*}
    \centering
    \includegraphics[width=1.0\linewidth]{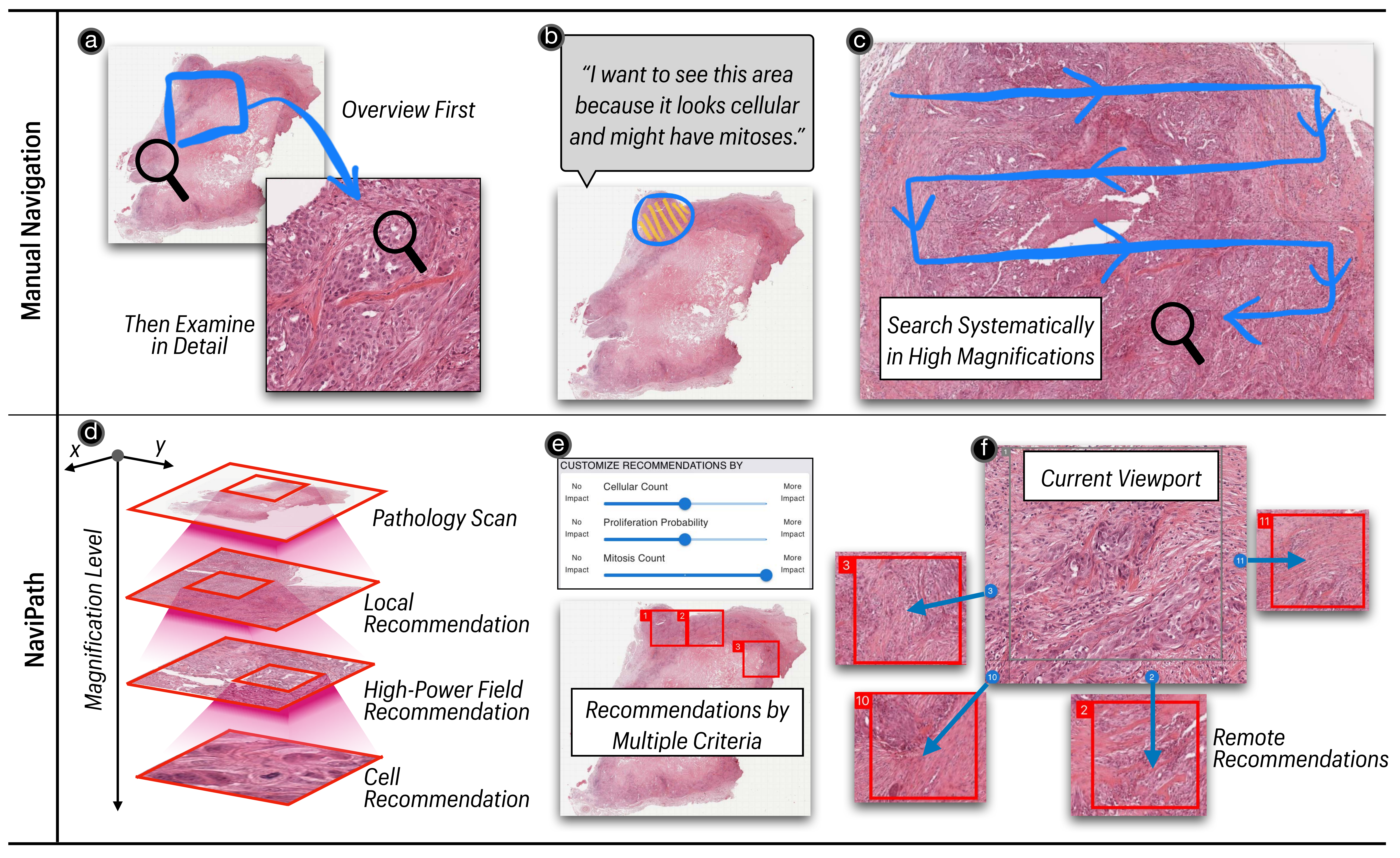}
    \caption{Comparison between pathologists' manual navigation in practice \vs \nap's designs. Observations on pathologists' manual navigation: (a) Pathologists usually overview a pathology scan with low magnifications, followed by switching to higher magnifications to examine regions of interest in detail; (b) Pathologists might refer to macroscopic patterns to locate ROIs in the low magnification; (c) Pathologists employ a systematical searching strategy in high magnifications. \nap's designs: (d) \nap~harnesses AI to generate hierarchical ``Local'', ``High-Power Field'', and ``Cell'' recommendations, covering multiple magnification levels; (e) \nap~utilizes AI to calculate three criteria that pathologists usually consider to generate recommendations; (f) Once in high magnifications, \nap~places navigation cues on the edge of the interface, enabling pathologists to jump to remote AI recommendations without manual panning.}
    \Description{A table that has two rows. The first row has three figures, showing how pathologists will navigate a high-resolution pathology scan manually. The second row has three figures, showing how human pathologists can use NaviPath to examine pathology scans.}
    \label{fig:nav_com}
\end{figure*}

\begin{enumerate}
    \item \textbf{Overview first, then detail}: Pathologists followed this pattern of interacting with visual data as found in earlier works \cite{545307, glueck2013model}: they started with an overview of the scan using low magnification, then selected a few \textbf{regions of interest (ROIs)} and studied each ROI in detail using higher magnifications (see Figure \ref{fig:nav_com}(a));
    \item \textbf{Using macroscopic patterns to locate ROIs in the low magnifications}: Pathologists referred to macroscopic patterns visible in low magnifications that were associated with occurrences of mitoses (see Figure \ref{fig:nav_com}(b)) to locate ROIs in low magnifications;
    \item \textbf{Low throughput in high magnifications}: Pathologists adopted a cautious and comprehensive navigation strategy (see Figure \ref{fig:nav_com}(c)) \cite{molin2015slide} to avoid missing crucial pathology patterns, causing low throughput under high magnifications.
\end{enumerate}

After accumulating the empirical evidence to verify existing knowledge in pathologists' navigation, we designed \nap~--- a human-AI collaborative navigation system that bridges the gap between AI and pathologists by integrating doctors' domain knowledge. Currently, we focus on pathologists' practices of examining mitosis as a showcase for \nap. Mirroring the three observations mentioned above, we propose three design components of \nap:

\begin{enumerate}
    \item \textbf{Hierarchical AI Recommendations}: As shown in Figure \ref{fig:nav_com}(d), \nap~employs AI to generate hierarchical recommendations across multiple magnification levels to support pathologists' ``overview first, then detail'' workflows. Specifically, the ``Local'' recommendation helps pathologists to quickly focus on a rough interest area in low magnification; the ``High-Power Field'' recommendation allows pathologists to narrow down and examine in detail using a median magnification level; and the ``Cell'' recommendation assists pathologists in adjudicating whether a suspected cell is mitotic in the highest magnification.
    \item \textbf{Customizable Recommendations by Multiple Criteria}: \nap~generates hierarchical AI recommendations with three criteria that pathologists usually consider to localize ROIs in practice (\ie cellular count, proliferation probability, and mitosis count). Furthermore, \nap~permits pathologists to customize AI recommendations according to their examination preferences by a group of slide-bars (Figure \ref{fig:nav_com}(e), top figure).
    \item  \textbf{Cue-Based Navigation for High Magnifications}: To cope with pathologists' low throughput under high magnifications, \nap~adapts the notion of existing cue-based navigation designs \cite{zellweger2003city} and places short-cut navigation cues on the edge of the viewport (Figure \ref{fig:nav_com}(f)). This design enables users to jump to remote AI recommendations without manual panning, which can improve pathologists' navigation efficiency.
\end{enumerate}

We recruited 15 medical professionals in pathology from five medical centers across two countries to validate \nap. We discovered that, compared to traditional manual navigation:

\begin{enumerate}
    \item Participants' navigation efficiencies were significantly improved ($p$=0.002, $r$=0.579, from Wilcoxon rank-sum test) with \nap: they saw more than twice the number of the target pathology pattern (\ie mitosis) in unit time on average;
    \item Both participants' precision and recall on identifying the target pathology pattern were significantly improved (precision: $p$<0.001, recall: $p$<0.001, from post-hoc Dunn's test) with \nap. Meanwhile, compared to the AI, participants' average recall and precision were improved by $20.21\%$ and $21.51\%$ by \nap, respectively;
    \item Participants reported significantly less mental effort ($p$<0.001, $r$=0.658, from Wilcoxon rank-sum test, same following), had higher confidence ($p$=0.004, $r$=0.530), and were more likely to use \nap~in the future ($p$=0.001, $r$=0.594), based on a post-study questionnaire.
\end{enumerate}

\subsection{Contributions}
We propose and validate the implementation of an AI-assisted tool in pathology ---~\nap~--- to enhance the navigation for pathologists by incorporating domain knowledge and considering workflow integration in practice. ~\nap~ could reduce pathologists' burdens by automating navigation with an AI-assisted algorithm while its collaborative workflow augments pathologists' work.
Throughout a user evaluation study with medical professionals, we demonstrated that our human + AI system could improve doctors' navigation efficiencies and lead to a higher examination quality. Instead of imposing an end-to-end, black-box AI into their workflows, this work closes the gap between medical professionals and AI by embedding doctors' domain knowledge and enabling them to delegate tasks to AI according to their preferences. Although majorly focused on mitosis in pathology, we further provide design insights for HCI researchers on how AI and medical professionals can work collaboratively to support medical decision-making in light of our observations in the evaluation study.



\section{Related Work}

This section introduces three domains of work related to \nap: \one interface designs to support pathologists' navigation, \two AI technologies for pathology, and \three human-AI collaboration to support medical decision-making.

\subsection{Supporting Pathologists' Navigation with Interface Designs}

Because the resolution of commercial off-the-shelf displays is significantly lower than pathology scans (up to $10^{12}$ pixels), intensive navigation is usually required for pathologists to search for features and make diagnoses \cite{ruddle2016design}. Since the issue roots in resolution differences, one intuitive solution is to introduce displays with larger physical sizes and resolutions to pathologists \cite{treanor2009virtual, randell2012using, wang2012surfaceslide, randell2015effect}. Literature has validated this solution, suggesting that pathologists utilized less pan and zoom interactions when using higher-resolution displays \cite{marchessoux2016}. However, improving hardware requires purchasing costly, bulky, and specialized devices. And we believe that interface designs that can aid pathologists to work with high resolution digital scans are more closely related to what NaviPath achieves.

A recent study suggests that employing appropriate interface designs can accelerate pathologists' examination processes comparable to those of using the optical microscope \cite{clarke2022faster}. Studies have well-explored designs to support navigating high-resolution images with limited size screens or displays \cite{bederson1996pad++, zellweger2003city, gustafson2008wedge, sarkar1992graphical, 10.1145/989863.989901}. Cockburn \etal~summarized these into four categories: \textbf{focus + context (F+C)}, \textbf{overview + detail (O+D)}, zooming, and cue-based \cite{cockburn2009review}.

In digital pathology, the main-stream open-source \cite{schneider2012nih, collins2007imagej, bankhead2017qupath, osd2022} and commercial \cite{imagescope2022} interfaces combine zooming and O+D designs, which include a zoomable canvas showing pathology scan details and an overview window that displays the thumbnail. Users can navigate high-resolution images with ``pan and zoom'' \cite{furnas1995space} interactions. However, criticisms suggest that such design demands a high mental effort and might be time-consuming \cite{jessup2021scope2screen, ruddle2016design}. To compensate for the limitation, Randell \etal~ improved the design by enlarging the overview to detail scale difference, enabling pathologists to pan more efficiently by moving the cursor in the `overview' window \cite{ruddle2016design}. Apart from O+D designs, Jessup \etal~ proposed an F+C interface for pathology image exploration \cite{jessup2021scope2screen}: a focal lens that magnifies the screen center and supports users' close-up examinations and explorations of multi-channeled pathology scans.

However, we argue that solely enhancing interface designs does not realize the full potential of digital scans. Because existing interface designs (without AI) lack support in assisting pathologists' visual searches \cite{palmer1995attention}, their navigation workflow can be substantially challenging while searching for small-sized, low-prevalence pathological patterns. Building upon the traditional O+D interface, the system proposed by this work adds AI and cue-based navigation, allowing pathologists to efficiently review AI findings with navigation cues.

\subsection{AI Technologies for Pathology}
Pathology has become an ``attractive target'' for applying AI because there exists a high variance in human diagnoses (\ie the problem of consistency) and a shortage of trained pathologists (\ie the issue of speed or efficiency) \cite{steiner2021closing}. Driven by high demand, the past decade has experienced a burst of publicly annotated datasets that cover a broad range of pathology practices, from conducting high-level diagnostic tasks (\eg identifying breast cancer metastasis \cite{litjens20181399}, classifying kidney transplant biopsies \cite{kers2022deep}) to seeing low-level histopathological patterns (\eg mitoses \cite{veta2019predicting, roux2013mitosis, aubreville2020completely, bertram2019large}). Following the enrichment of datasets, numerous works have proposed deep learning models to perform pathology image analysis, with some achieving in-lab performance comparable to human pathologists \cite{bejnordi2017diagnostic, campanella2019clinical}. Furthermore, multiple works have applied deep learning models for mitosis detection, which include \textbf{Convolution Neural Networks (CNNs)} \cite{tellez2018whole, gu2022detecting}, detection models (\eg RetinaNet \cite{aubreville2020completely}), or a combination of both \cite{mahmood2020artificial, li2018deepmitosis}. However, unlike humans, research has indicated that current deep learning models have a generalizability limitation --- their performance would deteriorate on the images with a domain shift (\eg a shift caused by a difference in the data handling procedure in medical centers) \cite{stacke2020measuring, aubreville2021quantifying}.

Setting aside the generalizability issue, the HCI problem of pathology using AI is its poor workflow integration: pathology is highly specialized domain in medicine, requiring specific expert knowledge and navigation strategies \cite{ruddle2016design, molin2015slide} to facilitate doctors' examination. As state-of-the-art AI focuses on pushing the performance with data-driven, `end-to-end' models, pathologists' needs for an AI's workflow integration is more or less ignored, which disincentives them from accepting and using AI in practice \cite{yang2016investigating}. In this work, instead of employing AI to replace pathologists, we adapt AI closely to doctors' domain knowledge of navigation, enabling them to work collaboratively with AI. Our validation study shows that our human + AI approach is recognized to have a better workflow integration and can help pathologists achieve higher precision and recall on average compared to start-of-the-art AI.

\subsection{Human-AI Collaboration for Medical Decision-Making}
Similar to how humans work with others, the human-AI collaboration envisions humans and machines working symbiotically \cite{licklider1960man} to achieve mutual goals \cite{wang2020human}. With the recent advancement of deep learning techniques, previous literature has established foundations of human-AI collaboration in the general domain (\eg design \cite{jeon2021fashionq} and content creation \cite{evirgen2022ganzilla}).  Furthermore, a number of HCI works have studied principles \cite{horvitz1999principles}, guidelines \cite{amershi2019guidelines}, design recommendations \cite{10.1145/3449084}, and information needs \cite{cai2019hello} to facilitate humans to work collaboratively with AI.

Following these pioneering works, research has investigated the broader applicability of human-AI collaboration for medical decision-making. For example, Beede \etal~ discovered socio-environmental factors that can influence AI performance, nurses workflows, and patient experiences while deploying a deep learning model to detect diabetic retinopathy \cite{beede2020human}. Wang \etal~ concluded the challenges of applying a clinical diagnostic support system in rural clinics \cite{wang2021brilliant}.  Lee \etal~ proposed a human-AI collaboration system for therapists' practices of rehabilitation assessments, and reported that the system can increase the consistency of decision-making \cite{lee2021human}. More recently, Fogliato \etal~ have studied the influence of human-AI workflows on veterinary radiologist readings of X-ray images, and revealed that doctors' findings were more aligned if AI suggestions were shown from the beginning \cite{fogliato2022goes}. Schaekermann \etal~discovered that implementing ambiguity-aware AI was more effective in guiding medical experts' attention to contentious portions while reviewing sheep EEG data, compared to conventional AI \cite{schaekermann2020ambiguity}. Calisto \etal~extended the designs of multi-modality radiology image viewing tools \cite{calisto2017towards, calisto2020breastscreening}. They built clinician-AI workflows for breast cancer image classification, suggesting that the human + AI approach could bring improvements in false-positives and false-negatives in diagnosis, user satisfaction, and time consumption \cite{calisto2021introduction, calisto2022breastscreening}.

Narrowing down to the pathology domain, promising works have employed a human + AI approach to support pathologists' examinations, bringing improvements in human errors \cite{wang2016deep, fdapaige2021}, between-subject agreements \cite{bulten2021artificial}, time consumption \cite{lindvall2021rapid}, and mental workload \cite{gu2021xpath}. For example, Lindvall \etal~ adapted the notion of \textbf{Rapid Serial Visual Presentation (RSVP)} \cite{spence2002rapid} and developed a rapid assisted visual search system, allowing pathologists to see and adjust the AI-generated ROIs by sensitivity \cite{lindvall2021rapid}. Gu \etal~ identified pathologists' challenges in practice and proposed a human-AI collaborative diagnosis system to perform multi-criteria, scan-level analysis for meningioma grading \cite{gu2021xpath}.  Notably, Cai \etal~ built a pathology \textbf{content-based image retrieval (CBIR)} system with an imperfect model --- pathologists could adjust the retrieved ROIs according to pathologist-defined concepts (\eg stroma) to cope with AI imperfections \cite{cai2019human}.

Extending the exciting success of human-AI collaborative systems in pathology, this work continues to explore user-centered, integrable designs to embed AI assistance into pathologists' navigation processes. Specifically, going beyond presenting AI results to inform pathologists \cite{cai2019human, gu2021xpath}, this work focuses on supporting the process with AI using designs that enable pathologists and AI to work symbiotically to navigate and gather information for diagnoses. Compared to previous human-AI navigation systems in pathology \cite{lindvall2021rapid}, \nap~incorporates the domain knowledge of pathologists' navigation, which can improve the workflow integration and better augment pathologists' routines of using AI as a companion.



\section{Formative Study \& System Requirements}

We conducted a formative study with six medical professionals in pathology (referred to as FP1 -- FP6) from two medical centers to study how pathologists examine digital scans for mitosis evaluation (see the supplementary material for the demographic information of participants). The participants were recruited using flyers sent in mailing lists and word-of-mouth. For each participant, we first introduced the mission of the project. Then, we presented a pathology scan selected from \cite{aubreville2020completely}, and asked participants to assess the activity of mitosis (a pathological pattern). We followed up with a semi-structured interview and inquired how they navigated the scan to find mitoses. Finally, we presented a series of candidate mock-ups of \nap~and collected participant feedback. The length of the semi-structured interview was about 30 minutes, and the average duration of each study was about 60 minutes.

\subsection{Observations}
We analyzed the transcribed interview recording using the following approach: first, two researchers summarized the observations individually; then, a third researcher reviewed the observations and addressed the disagreements. We concluded three observations of how pathologists navigate pathology scans (without AI) in their practice, which cross-validated findings from previous work on humans' navigation patterns in high-dimensional visual data.

\begin{itemize}
    \item \textbf{O1: Overview first, then detail.}
    To search for mitoses, pathologists would first stay in low magnifications to get an overview of the scan, then select a few ROIs and study each ROI in greater detail using higher magnifications. Such a routine was also described in previous works in the general domain of information searching \cite{545307, glueck2013model} and pathology \cite{ruddle2016design}. Pathologists adapted the searching strategy because of the size difference between mitoses and pathology scans --- mitosis is a small-sized pathology feature and can hardly be observed without high magnifications (\ie $\sim$ $\times$400 magnification). However, scanning the entire slide systematically in $\times$400 \cite{molin2015slide} can be substantially time-consuming because the field of view under $\times$400 is small compared to the pathology scan: a field of view under $\times$400 has a size of $0.16mm^2$, while a typical $\times$400 pathology scan usually has a size of $\sim 100 mm^2$. In our study, all six participants searched for mitoses more efficiently: first, they rapidly covered the scan in low magnifications (<$\times$50) as an overview. After that, they selected a few ROIs to proceed: for each ROI, they switched to medium-magnification ($\sim \times$200) to maximize their fields-of-view while preserving cellular details. If a suspected cell was found, they would dive into high-magnification ($\times$400) and make an adjudication.
    
    \item \textbf{O2: Using macroscopic patterns to locate ROIs in the low magnification.}
    To locate the mitosis, pathologists used not only the microscopic features (only visible in $\times$400) but also referred to macroscopic patterns (visible even in <$\times$50) that were associated with the occurrences of mitoses. Specifically, pathologists located ROIs in low-magnification by evaluating the cell density --- \textit{``if it (an ROI) is more cellular, it is more likely to have mitoses''}(FP3). 
    
    \item \textbf{O3: Low throughput in higher magnifications.} While pathologists relied on the cell density to select ROIs from low magnifications, they were likely to `get lost' once they had switched to higher magnifications. This is because there was a lack of visual landmarks under high magnifications in tumor scans (\ie the `desert fog' problem \cite{jul1998critical}). From the study, we observed that some participants preferred to use a cautious and comprehensive navigation strategy \cite{molin2015slide} (see Figure \ref{fig:nav_com}(c)) to avoid missing critical findings that might overturn the diagnosis. However, because not all areas under the high magnifications include mitoses, the navigation strategy might be less efficient and more prone to causing fatigue.
\end{itemize}

\subsection{System Requirements}
Based on our observations, we propose the following three system requirements for human-AI navigation systems for pathologists:

\begin{itemize}
    \item \textbf{R1: Covering multiple magnification levels}. In accordance with pathologists' ``overview first, then detail'' navigation processes, the system should provide AI support across multiple magnification levels. For example, recommendations in low magnifications can draw pathologists' attention by pointing out rough areas of interest, while those in higher magnifications should offer more precise guidance in locating ROIs.
    
    \item \textbf{R2: Incorporating pathologists' domain knowledge}. To bridge the gap between pathologists and AI, instead of employing end-to-end, black-box AI, the system should adapt AI closely to pathologists' domain knowledge and involve criteria that pathologists use in practice to generate AI recommendations. Moreover, because pathologists might have diverse preferences and AI can be imperfect \cite{stacke2020measuring, aubreville2021quantifying}, the system should allow users to customize AI recommendations by emphasizing or ruling-out specific criteria.
    
    \item \textbf{R3: Accelerating navigation in high magnifications}. To address the low-throughput issue, the system should offer interface designs that enable users to navigate efficiently among the AI recommendations in high magnifications, without getting lost.
    
\end{itemize}



\section{Design of NaviPath}

In this section, we first introduce four design components used in \nap. We then describe how \nap~augments pathologists' navigation by describing an example workflow.

\subsection{Design Components}
Corresponding to the three system requirements, we propose three key designs in \nap: \textbf{Hierarchical AI Recommendations}, \textbf{Customizable Recommendations by Multiple Criteria}, and \textbf{Cue-Based Navigation for High Magnifications}. Furthermore, we employ the design of \textbf{Explaining Each Recommendation} to help pathologists comprehend AI findings.

\subsubsection{Hierarchical AI Recommendations} 
\begin{figure*}
    \centering
    \includegraphics[width=0.9\linewidth]{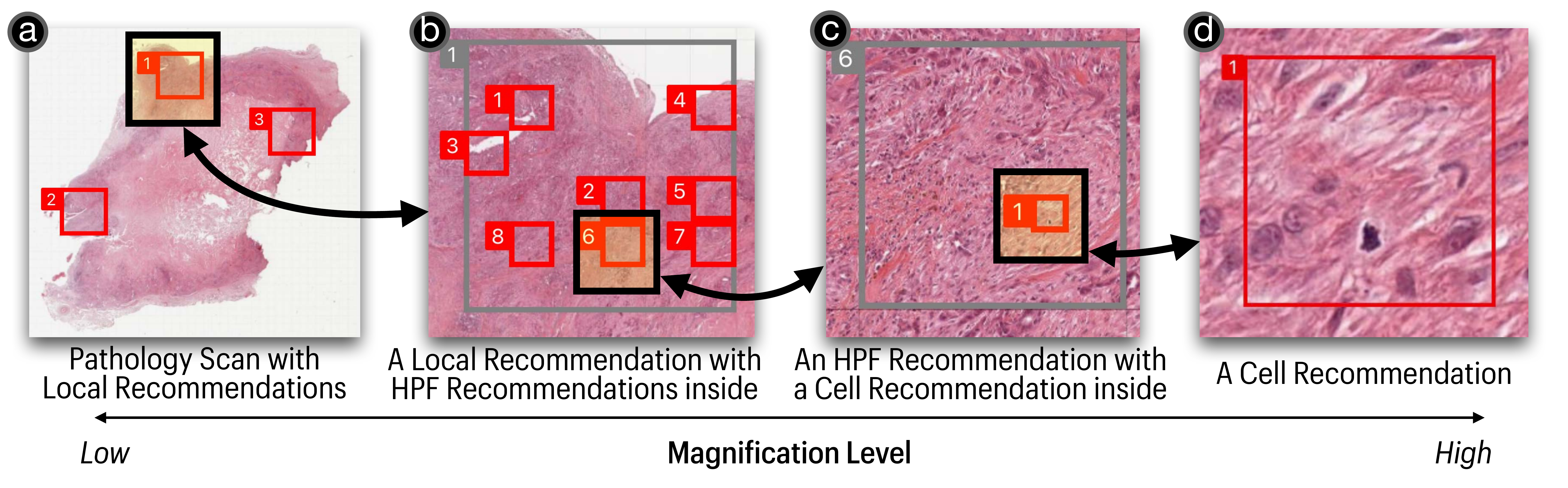}
    \caption{\nap~generates hierarchical AI recommendations across multiple magnification levels: (a) Local recommendations (red boxes) lie in the lowest magnification, and can be seen directly on the pathology scan without zooming; (b) there are multiple High-Power Field (HPF) recommendations (red boxes) inside one Local recommendation (gray box); (c) once in an HPF recommendation (the gray box), users can select and see (d) a Cell recommendation with the highest magnification.}
    \Description{Four subfigures are shown in a row. Each of the subfigrue represents NaviPath's AI recommendations on each magnification level. From left to right, these subfigures show NaviPath's recommendation boxes from the local recommendation (with the largest size) in the low magnification, HPF recommendation (with the medium size), in the medium magnification, to the Cell recommendation (with the smallest size) in the high magnification. The last figure illustrates that a mitosis figure can be seen in a Cell recommendation.}
    \label{fig:nap_rec_design}
\end{figure*}

Following pathologists' navigation processes for mitosis searching, \nap~offers AI recommendations of three sizes\footnote{Specific sizes were justified by consulting with a board-certified pathologist (experience = 10 years)} to provide assistance across multiple magnification levels (system requirement \textbf{R1}): 

\begin{enumerate}
    \item The ``Local'' recommendation (size=10,080$\times$10,080 pixels\footnote{The size of one pixel is 0.25$\mu$m.}) simulates pathologists' overviewing processes in low magnification. As shown in Figure \ref{fig:nap_rec_design}(a), the recommendations are red boxes visible in the pathology scan without zooming. Local recommendations can provide rough directional guidance for pathologists; users can prioritize their examination on AI-selected regions without evaluating the scan manually.
    \item There are multiple \textbf{``High-Power Field'' (HPF)} recommendations (size=1,680$\times$1,680 pixels) within a Local recommendation (Figure \ref{fig:nap_rec_design}(b), red boxes). The HPF recommendation gives more precise ROIs at a higher magnification level, allowing users to examine them in detail. It has the same field of view as $\times$400 in optical microscopes that pathologists use in practice, freeing them from spending extra effort on adapting to the digital interface.
    \item The ``Cell'' recommendation (size=240$\times$240, Figure \ref{fig:nap_rec_design}(d)) points out the most precise location of each suspected mitosis reported by AI. It augments pathologists' mitosis evaluations by transforming a visual search task (\ie finding where mitoses are) into the adjudication (\ie whether a Cell recommendation includes mitosis).
\end{enumerate}

For all three levels, users can select a recommendation by double-clicking on it, and \nap~ will automatically zoom and center the viewport to the selected recommendation. Therefore, with hierarchical AI recommendations, users can proceed through magnification levels by selecting recommendations on the next level (\eg Figure \ref{fig:nap_rec_design}(a)$\rightarrow$(b), (b)$\rightarrow$(c), (c)$\rightarrow$(d)). Users may ignore the recommendation if an undesired one appears.

\subsubsection{Customizable Recommendations by Multiple Criteria}
\begin{figure*}
    \centering
    \includegraphics[width=1.0\linewidth]{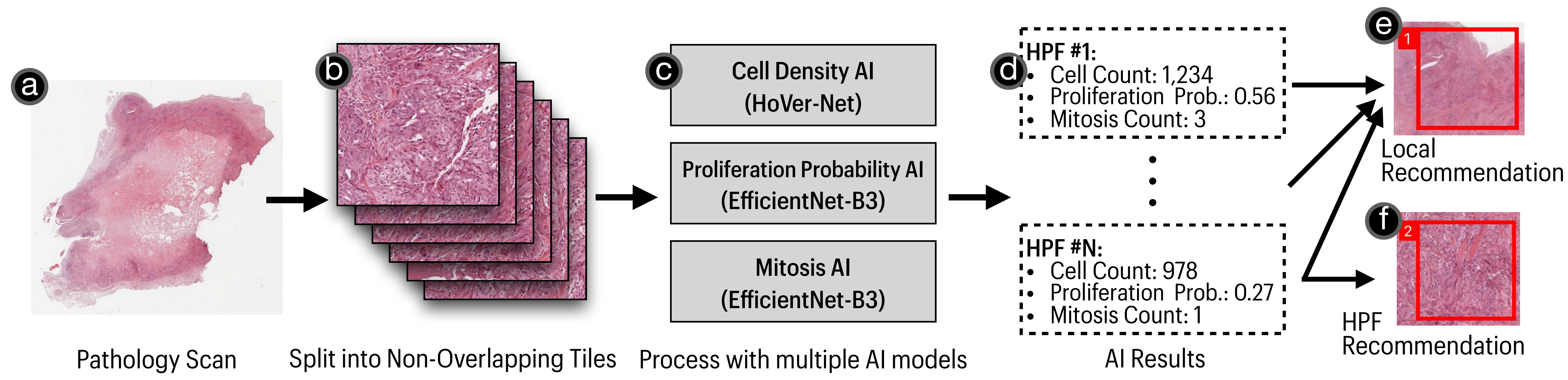}
    \caption{Generating Local and HPF recommendations with multiple criteria: (a) a pathology scan is first (b) split into non-overlapping tiles. Then, \nap~uses (c) three AI models to analyze each tile to obtain (d) scores of cellular count, proliferation probability, and mitosis count. \nap~will (e) aggregate scores from multiple tiles to generate Local recommendations, or (f) directly use these scores for HPF recommendations.}
    \Description{This figure shows a flowchart that describes NaviPath's data processing pipeline. Form left to right, there are five parts: (1) the left most subfigure illustrates a pathology scan; (2) the second subfigure tells that the pathology scan is split into a deck of tiles with smaller dimensions; (3) the third subfigure describes that these tiles were processed by three AI models; (4) the fourth subfigure gives examples of the results in each tile processed by AI; (5) the fifth subfigure shows two recommendations of NaviPath was generated according to the AI results in each tile.}
    \label{fig:nap-ai}
\end{figure*}

NaviPath embeds pathologists' domain knowledge and employs three deep learning models (Figure \ref{fig:nap-ai}(c)) to calculate three criteria for obtaining Local and HPF recommendations (system requirement \textbf{R2}):
\begin{enumerate}
    \item \textbf{Cellular Count}: Similar to how pathologists leverage the cell density to locate ROIs in the low magnification, \nap~ employs a state-of-the-art nuclei segmentation model (\ie HoVer-Net) to count cell numbers and capture cellular areas from the pathology scan.
    \item \textbf{Proliferation Probability}: Mimicking pathologists' judgements of whether an area needs further attention in $\times$400 from $\times$200 views, \nap~uses an EfficientNet-b3 model \cite{tan2019efficientnet} to predict the proliferation probability --- a criterion that relates to whether an ROI is likely to include mitosis, based on AI's impressions from $\times$200 magnification.
    \item \textbf{Mitosis Count}: Corresponding to pathologists' mitoses searching in $\times$400, \nap~utilizes a classification model (\ie EfficientNet-b3) to detect mitotic figures from the highest magnification.
\end{enumerate}

As for Cell recommendations, \nap~directly pulls the positive results from the mitosis AI and visualizes them on the interface.

\begin{figure*}
    \centering
    \includegraphics[width=1.0\linewidth]{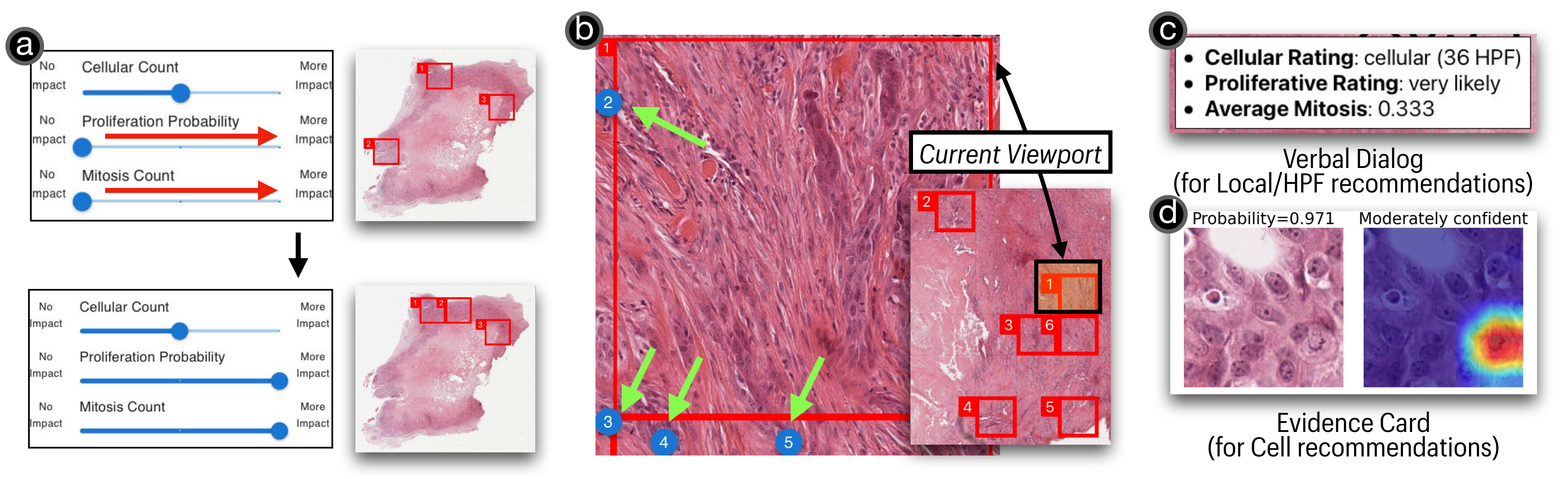}
    \caption{(a) \nap~supports users to customize AI recommendations with a group of slide-bars: users can emphasize or rule out each of the three criteria (\ie cellular count, proliferation probability, mitosis count) for \nap's recommendations; (b) \nap~places navigation cues (pointed by arrows) that enable users to hop to remote recommendations. The figure on the right provides an overview of off-screen recommendations; (c) An example of \nap's verbal dialog explanation for Local/HPF recommendations; (d) An example of the explanation card for \nap's Cell recommendations.}
    \label{fig:nap-cus}
    \Description{This figure shows three subfigures that describe NaviPath's functionality. (1) the subfigure illustrates NaviPath's slide-bars, and users can adjust the slide-bar to customize NaviPath's recommendation; (2) the subfigure shows NaviPath's navigation cues, a number of small buttons on the edge of NaviPath's interface, where users can jump to a remote area with one click; (3) the subfigure demonstrates examples of NaviPath's AI recommendations, including an example verbal dialog for local/HPF recommendation, and an example of evidence card for Cell recommendation. The text in the verbal dialog says "Cellular rating: cellular (36HPF), Proliferative Rating: "very likely", Average Mitosis: 0.333. The evidence card shows a cell recommendation and a heatmap, the text on the evidence card says: "Probability = 0.97, Moderately confident"}
\end{figure*}

\begin{figure*}
    \centering
    \includegraphics[width=1.0\linewidth]{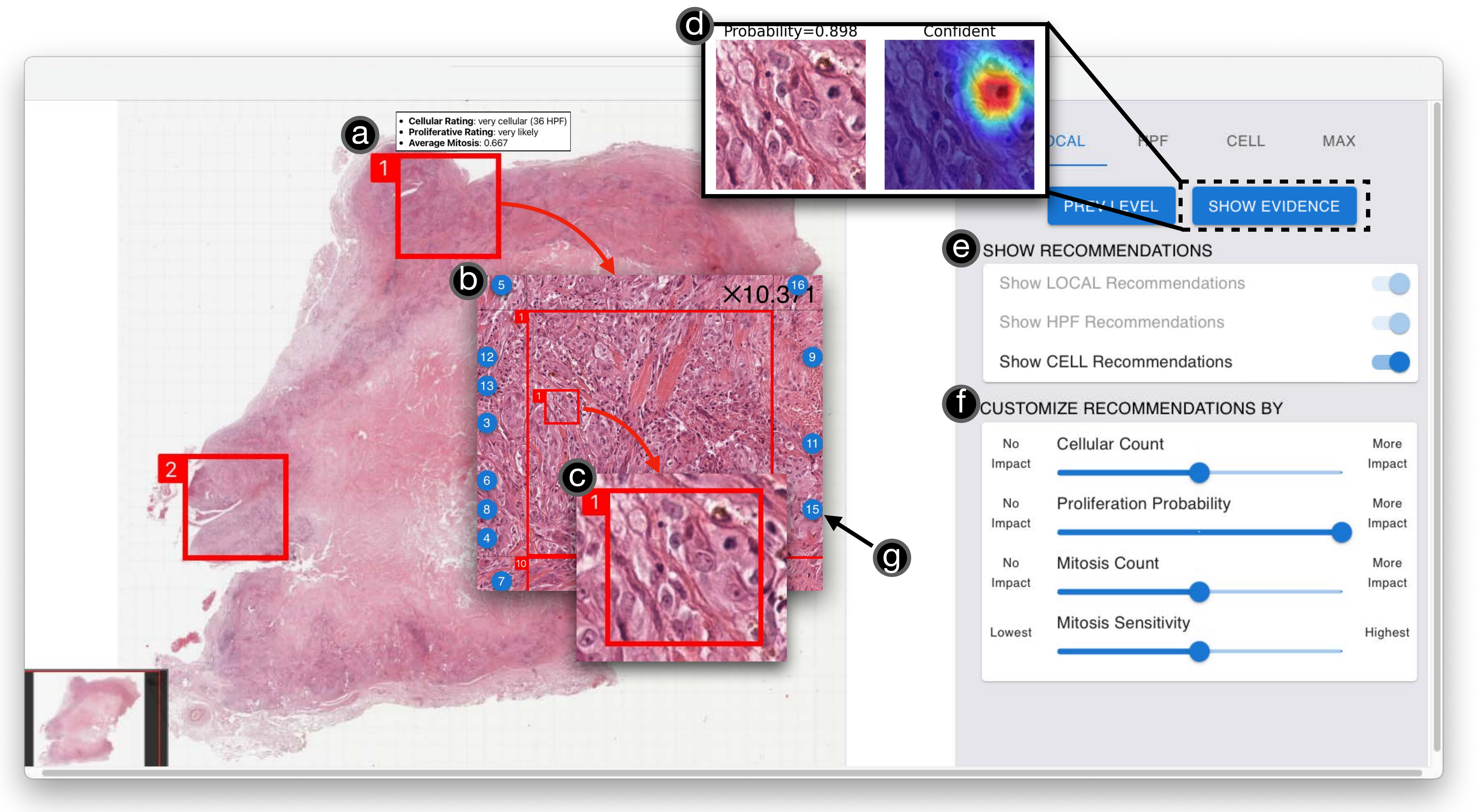}
    \caption{Overview of \nap's interface. (a) A Local recommendation (red box) with an explanation dialog. The number on the top-left corner represents the index of the recommendation (same for HPF and Cell recommendations); (b) An example of an HPF recommendation; (c) An example of a Cell recommendation; (d) An explanation card for a Cell recommendation, including the AI probability, confidence level, and a saliency map; (e) Users can switch on and see each level of recommendations on-demand; (f) Users can customize the recommendations with a group of slide-bars; (g) A navigation cue that allows users to jump to a remote recommendation. The number indicates the index of the remote recommendation.}
    \label{fig:nap_ui}
    \Description{The figure illustrates NaviPath's interface. The left side of the interface is a zoomable canvas that shows pathology scan, where NaviPath's AI recommendation boxes with verbal dialog explanations are displayed. The right side provides a list of AI features: from top to bottom: (1) there are two buttons "PREV LEVEL" and "SHOW EVIDENCE". "PREV LEVEL" can help users to go back to the previous level of recommendations. "SHOW EVIDENCE" can help users see the evidence card for the Cell recommendation; (2) below the two buttons are three switches that allows users to can turn on/off Local, HPF, Cell recommendations respectively; (3) four sliders under the text of "CUSTOMIZE RECOMMENDATIONS BY" were shown in the bottom. Users can slide each slide-bar left/right to increase/decrease the impact of the criterion "cellular count", "proliferation probability", "mitosis count", and "mitosis sensitivity" individually. NaviPath's AI recommendations will be updated according to users' slide-bar status.}
\end{figure*}

Since pathologists might use the three criteria differently in practice, \nap~supports users to customize AI recommendations by emphasizing or ruling out specific criteria with a group of slide-bars, as shown in Figure \ref{fig:nap-cus}(a). For example, giving the ``Proliferation Probability'' and ``Mitosis Count'' higher weight by moving the slide-bar to the right will force \nap's recommendations to lean on these criteria. \nap~will then re-calculate and update recommendations based on the user's input. What's more, users can also adjust the sensitivity of recommendations. For example, if users wish to see more recommendations, they could tune up the ``Mitosis Sensitivity'' slide-bar (see Figure \ref{fig:nap_ui}(f), the fourth slide-bar). 

\nap~ranks all recommendations according to the current customization setting.  Based on the ranking result, it assigns each AI recommendation an index (\eg Figure \ref{fig:nap_ui}(a), the number on the top-left corner of the recommendation). The smaller the index, the  greater the importance and need to be examined with high priority. The index number gives users ``actionable'' advice \cite{10.1145/3449084} and can help them focus on critical areas in limited time. Please refer to the supplementary material for the implementation details of AI models and the recommendation ranking algorithm.

\subsubsection{Improving Navigation in High Magnifications} Following system requirement \textbf{R3}, \nap~uses two designs to optimize pathologists' navigation in high magnifications:

First, \nap~enables pathologists to pan discretely in high magnifications. Specifically, after examining each HPF recommendation, users can double-click on the screen's edge to pan discretely to an adjacent one. Compared to the conventional manual panning with mouse-dragging, this design can accelerate users' interaction speeds: according to Fitt's Law \cite{fitts1954information}, screen edges have infinite width, so it follows that.

Moreover, to increase pathologists' efficiency in seeing remote recommendations, \nap~adapts the notion of citylight \cite{zellweger2003city} by placing navigation cues on the edge of the interface (Figure \ref{fig:nap-cus}(b), pointed by arrows). The location of the navigation cue indicates the relative direction between the remote HPF recommendation and the current viewport, while the number represents the ranked index of each recommendation. With navigation cues, users can become aware of the spatial distribution and importance of off-screen targets. They can also click on navigation cues to hop to remote HPF recommendations without manual panning.

\subsubsection{Explaining Each Recommendation}
Since one criticism of deep learning models in pathology is that there is a lack of interpretability \cite{stiglic2020interpretability}, \textbf{explainable AI (XAI)} techniques have been utilized to make AI ``transparent, understandable and reliable'' to pathologist users \cite{pocevivciute2020survey}. In \nap, we followed the suggestions from \cite{10.1145/3449084} and attached an explanation for each AI recommendation. Specifically, for Local and HPF recommendations, \nap~presents users with a verbal dialog, which includes qualitative descriptions of AI results on the cellular count, proliferation probability, and mitosis count (Figure \ref{fig:nap-cus}(c)). The dialog helps users decide whether they should select and study recommended areas. Moreover, \nap~adapts the design of a previous human-AI pathology system \cite{gu2021xpath} and explains each Cell recommendation with an explanation card (Figure \ref{fig:nap-cus}(d)). The explanation card demonstrates the classification probability, the confidence level, and a saliency map for a positive mitosis classification result, which provides information from AI's perspective to assist pathologists' mitosis adjudications. Detailed procedures of explanation generation are described in the supplementary material.

\subsection{Navigating with \nap}
A typical page of \nap~is shown in Figure \ref{fig:nap_ui}. A user's workflow in \nap~starts by switching on (Figure \ref{fig:nap_ui}e) and seeing Local recommendations (Figure \ref{fig:nap_ui}a). The number on the top-left corner of each recommendation box is the ranking index, and users may view recommendations by ascending index order. In each Local recommendation, users can continue to drill down and see HPF recommendations (Figure \ref{fig:nap_ui}b). In each HPF recommendation, users can continue to see Cell recommendations (Figure \ref{fig:nap_ui}c) that show the precise locations of detected mitoses. For each Cell recommendation, users can view an explanation card on-demand (Figure \ref{fig:nap_ui}d). After examining each HPF recommendation, users may click on the numbered navigation cue (Figure \ref{fig:nap_ui}g) to hop to a remote HPF recommendation. During users' examination, they may customize the recommendations by interacting with a group of slide-bars (Figure \ref{fig:nap_ui}f). Users’ workflow ends when they are confident of signing out the case.

\section{Technical Evaluation}

We conducted a technical validation study and reported the performance of the three AI models in \nap. Specifically, we applied classification models for mitosis and proliferation probability on the eight test scans selected from \cite{aubreville2020completely}. We cross-referenced the AI results and ground-truth labels to calculate F1 scores. The ground-truth labels for mitosis detection and proliferation probability calculation were acquired/generated from the annotations provided in \cite{aubreville2020completely}. For the cellular count calculation, we applied the model to 50 randomly-picked areas (size=$512\times 512$ pixels under $\times$400 magnification) from pathology scans. Then we compared the AI result with the cellular count reported by a graduate student, who had been briefly instructed by a pathologist (experience = 10 years).

The results showed that the mitosis detection model achieved an F1 score of 0.673 (precision: 0.703, recall: 0.650) when using a probability threshold of 0.85. The F1 score for the proliferation probability model was 0.472 (precision: 0.544, recall: 0.416, probability threshold: 0.77). The average error rate of the cell counting model was 14.95\%.

Although we tried to train the model for mitosis detection following a recent work \cite{gu2022detecting}, the performance of the mitosis AI was still not perfect: tuning down the threshold and setting the recall as 0.85 caused the precision score to drop to 0.216. That is, the number of false-positive instances would have been 3.62$\times$ the true-positive ones. The proliferation probability model performance was also not satisfactory, likely due to the misalignment in label distribution between train/validation and test sets: while 15.0\% of train/validation data were positive, only 4.7\% of test data were positive.



\section{Work Sessions with Pathologists}
We conducted work sessions with medical professionals in pathology to validate \nap, studying three research questions:

\begin{itemize}
    \item \textbf{RQ1}: Can \nap~(as a human + AI approach) increase pathologists' precision and recall in identifying the pathological features (in this case, mitosis)? 
    
    \item \textbf{RQ2}: Can \nap~ save pathologists time and effort?
    
    \item \textbf{RQ3}: Compared to manual navigation, what is the benefit of using \nap?
\end{itemize}

We designed three testing conditions to support the system validation on the three \textbf{RQ}s:

\begin{itemize}
    \item \textbf{C1 (Human Only)}: Participants navigate a pathology scan viewer without any AI assistance;
    \item \textbf{C2 (Human + AI)}: Participants navigate the pathology scan with \nap;
    \item \textbf{C3 (AI Only)}: AI-automatic reporting without humans; 
\end{itemize}

\subsection{Participants}

We recruited 15 medical professionals in pathology from five medical centers across two countries, including 13 residents, one fellow (P7), and one attending (P15). The participants were recruited through flyers sent in mailing lists and word-of-mouth. The demographic information of the participants is shown in Table \ref{user_dem}. All participants had received at least two years of pathology residency training to be qualified for the study (average experience $\mu$=3.47 years, $Std$=0.88 years). 14/15 participants had experience in seeing pathology scans before the study (daily: 3, weekly: 6, bi-weekly: 3, monthly: 1, within one year:1). The primary purpose for using pathology scans was for learning, and the most mentioned digital pathology interface was Aperio Imagescope \cite{imagescope2022}.

\subsection{Data \& Apparatus}
We collected eight pathology scans of canine mammary carcinoma from a public dataset \cite{aubreville2020completely}. The average size of these scans was 7.15 giga-pixels. We acquired the ground-truth mitosis annotations from the same dataset \cite{aubreville2020completely}. Overall, the average \textbf{mitotic rate} (\ie \textbf{MR}, mitotic count per unit area\footnote{\url{https://www.cancer.gov/publications/dictionaries/cancer-terms/def/mitotic-rate}}) was 1.022/mm$^2$ (0.164/HPF). We selected two scans for tutorial purposes, leaving the other six for testing (Scan 1-6 in Table \ref{user_dem}). To generate AI detections, the scans were pre-processed with a local server with a 24-core CPU, 64 GB memory, and an Nvidia RTX-3090 graphics card. After that, we loaded the pre-processed results into \nap~(\textbf{C2}). For a comparison, we developed a baseline pathology scan viewer with a basic O+D design, where pathologists were required to navigate manually to evaluate mitosis activity (\textbf{C1}). During the study, we referred to the manual baseline system as `\textbf{system 1}' and \nap~ as `\textbf{system 2}' to avoid bias.

\begin{table*}
\centering
\caption{Demographic information \& arrangements of the participants in the work sessions. The number `1' indicates that the scan was examined with system 1 (baseline manual system), while `2' was with system 2 (\nap). MC1-3 are located in one country, and MC4-5 are in another.}
\scalebox{0.95}{
\begin{tabular}{c | c c c | c c c c c c c}
\toprule
 \textbf{ID} & \makecell{\textbf{Years of} \\ \textbf{Experience}}  & \makecell{\textbf{Frequency of Seeing} \\ \textbf{ Pathology Scans}} & \makecell{\textbf{Medical} \\ \textbf{Center}}& \textbf{Scan 1} & \textbf{Scan 2} & \textbf{Scan 3} & \textbf{Scan 4} & \textbf{Scan 5} & \textbf{Scan 6}\\
\midrule
 P1 & 4 & Weekly & MC1 & 2 & & & & 1 & \\
 P2 & 3 & Never & MC2 & 1 & & & & 2 &\\
 P3 & 4 & Bi-Weekly & MC3 & 2 & & & & 1 & \\
 P4 & 4 & Weekly & MC3 & 1 & & & & 2 & \\
 P5 & 3  & Daily & MC4 & & & & 2 & & 1\\
 P6 & 2  & Weekly & MC1 & & & 1 & 2 & &\\
 P7 & 5  & Daily & MC3 & & & 1 & 2 & &\\
 P8 & 4  & Bi-Weekly & MC3 & & & 2 & 1 & &\\
 P9 & 4  & Daily & MC3 & & 1 & & & & 2 \\
 P10 & 3  & Weekly & MC4 && 1 & & & & 2 \\
 P11 & 2  & Bi-Weekly & MC4 & & 2 & & & & 1 \\
 P12 & 3  & Weekly & MC4 && 2 & & & & 1 \\ 
 P13 & 3  & Monthly & MC4 && & 1 & & & 2 \\ 
 P14 & 3  & Within One Year & MC4 && & 2& 1& & \\ 
 P15 & 5  & Weekly & MC5 && 2 & 1& & & \\
\bottomrule
\end{tabular}}
\label{user_dem}
\end{table*}

\subsection{Task \& procedure}
All sessions were conducted online over Zoom. Participants were first shown a tutorial video ($\sim$10 minutes) of the manual baseline system and \nap. After they had watched the video, they were given links to both systems, which were accessed through the web browser. Next, each participant was instructed to perform a pathology task of assessing the mitotic activity of one pathology scan using system 1/system 2, and another with system 2/system 1. During the formative study, we discovered that pathologists might memorize the hot-spot areas of a pathology scan that they had examined before by recognizing tumor contours, even after several months. Therefore, instead of letting a participant see the same scan after a wash-out period, we instructed participants to read different scans in the work sessions (see Table \ref{user_dem}). The order of seeing the scans in each session was counterbalanced across participants. During each session, participants were required to evaluate the mitotic activity following the College of American Pathologists (CAP) cancer protocol\footnote{\url{https://documents.cap.org/protocols/cp-cns-18protocol-4000.pdf}.}, which is similar to how pathologists examine the scan in practice. Finally, participants entered a post-study structured interview that included a set of Likert questions and short answers. The average duration of each study was about 65 minutes.

\subsection{Measurements}
We collected three sources of responses from users during the work session: first, we recorded participants' interactions with both systems. Second, after they had finished examining each scan, we saved participants' reportings of mitoses. Third, from the final interview, we collected participants' responses to the questionnaire. Following previous HCI research on pathology navigation \cite{ruddle2016design} and pathology AI \cite{cai2019human}, we investigated the research questions with the following measurements:

For \textbf{RQ1}, we obtained the participants' mitosis reportings with the baseline \textbf{C1}, \nap~(\textbf{C2}), and AI (\textbf{C3}). We then cross-referenced them with ground-truth mitosis labels and calculated precision and recall scores. Because each participant may visit different ROIs in each trial, we individually calculated the AI's precision and recall scores (\textbf{C3}) within the areas visited by each participant in \textbf{C2}. Therefore, we can study whether the improvements in \textbf{C2} are brought by NaviPath’s AI or its human-AI workflow.

For \textbf{RQ2}, we first calculated participants' average time cost on each scan. We also evaluated each participant's navigation efficiencies by counting the number of \textit{ground truth mitosis} within the areas visited by participants in each trial and divided it by the time length. After that, we averaged the results across the participants for \textbf{C1} and \textbf{C2} individually. Here, we did not count the \textit{mitosis reported by participants} as in \textbf{RQ1} to rule out the difference in participants' capabilities in locating mitoses. Finally, to evaluate the cognitive workload of using both systems, we asked the participants to answer two seven-scaled Likert NASA TLX questions (\ie mental demand and frustration dimensions, Table \ref{tab:question_response} Q1, Q2)) \cite{HART1988139}.

For \textbf{RQ3}, we first analyzed the interaction logs and summarized participants' interaction frequencies with both systems (\ie zoom, pan, selecting recommendations). What's more, we inquired about participants' ratings on system's capabilities for mitosis searching (Table \ref{tab:question_response} Q3), their confidence in the mitosis reportings (Table \ref{tab:question_response} Q4), attitudes toward using the system in the future (Table \ref{tab:question_response} Q5), and overall preference of system 1 \vs system 2 (Table \ref{tab:question_response} Q6).

Last but not least, to figure out whether each \nap~component is useful for pathologists, we asked the participants to rate each component (Figure \ref{fig:util-rating}) with a seven-scaled Likert question: \one \textit{``Is this feature useful to your examination?''} (1= Not useful at all $\rightarrow$ 7=Very useful); \two \textit{``Compared to System 1, does this feature require extra effort?''}(1=No effort at all $\rightarrow$ 7=A lot of effort).



\section{Result \& Findings}

In this section, we first answer our initial research questions based on the information collected from work sessions. We then summarize the qualitative findings on pathologists' navigation traces.

\subsection{Results for Research Questions}

\begin{figure*}
    \centering
    \includegraphics[width=0.8\linewidth]{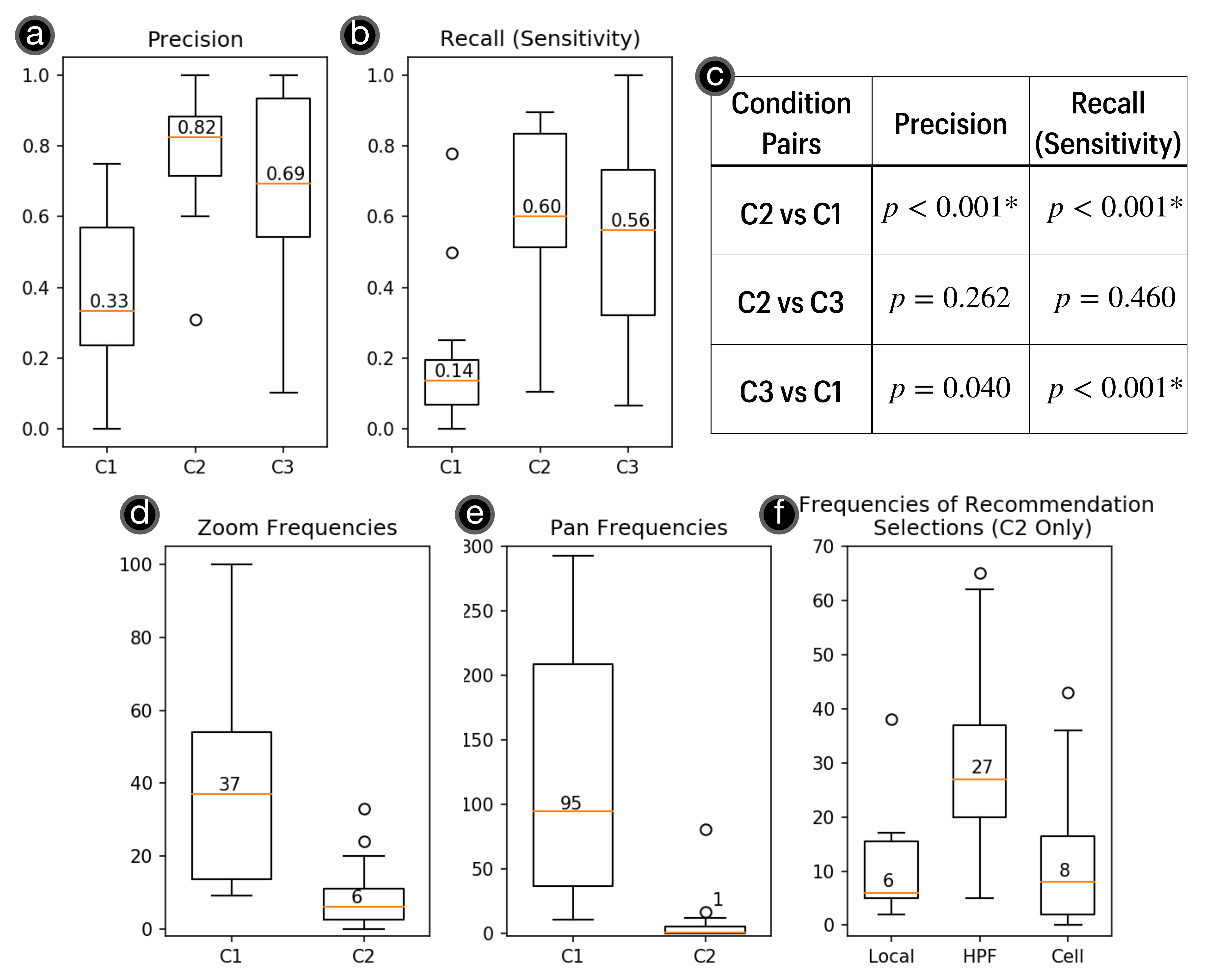}
    \caption{Boxplot visualizations of the (a) precision and (b) recall (sensitivity) from mitosis reportings under the conditions of \textbf{C1}, \textbf{C2}, and \textbf{C3}. The colored lines and the figures above indicate the median values of each condition. The dots are the outliers. (c) The results of pair-wise significance comparison among \textbf{C1}, \textbf{C2}, and \textbf{C3} using a post-hoc Dunn's test with Bonferroni correction ($\alpha$=0.05). The values marked with $^*$ indicates that the Null hypothesis can be rejected because the $p<\alpha/2$. (d) Participants' zoom interaction frequencies under \textbf{C1} and \textbf{C2}. (e) Participants' pan interaction frequencies under \textbf{C1} and \textbf{C2}; (c) Frequencies of participants' selecting Local, HPF, and Cell recommendations under \textbf{C2}. Note that one participant might select the same recommendation multiple times in each trial.}
    \label{fig:prec-recall}
    \Description{Five subfigures were shown. (a) Boxplot of participants' precision. C2's box is higher than C3's, and C3's is higher than C1. The medium scores of C1, C2, and C3 are 0.33, 0.82, 0.69. (b) Boxplot of participants' recall. C2's box is slightly higher than C3's, and C3's is higher than C1. The medium scores of C1, C2, and C3 are 0.14, 0.60, 0.56. (c) A table that shows pair-wise significance comparison p values of conditions C1, C2, C3's precision and recall. C2 vs. C1's precision p<0.001, recall p<0.001. C2 vs. C3's precision p=0.262, recall p=0.460. C3 vs. C1 precision=0.040, recall p<0.001. (d) Boxplot of participants' zoom interaction frequencies with C1 and C2. C1's box is much higher than C2's. The medium scores of C1 and C2 are 33 and 6. (e) Boxplot of participants' pan interaction frequencies with C1 and C2. C1's box is much higher than C2's. The medium scores of C1 and C2 are 95 and 1. (f) Boxplot of participants' Local, HPF, and Cell recommendation selection frequencies with C2. HPF's box is much higher than C1's and C3's. C1's and C3's boxes have similar height. The medium scores of Local, HPF, and Cell are 6, 27, 8. }
\end{figure*}

\subsubsection{RQ1:  Can \nap~increase pathologists' precision and recall in identifying the pathological features? }

We calculated the precision and recall (sensitivity) of participants' mitosis reportings with manual navigation (\textbf{C1}), \nap~(\textbf{C2}), and AI-automated reportings (\textbf{C3}) (Figure \ref{fig:prec-recall}(a)-(b)). The median precision under \textbf{C1}, \textbf{C2}, and \textbf{C3} were 0.33, 0.82, and 0.69, respectively (average $\mu$=0.40, 0.78, 0.64, standard deviation $Std$=0.22, 0.17, 0.31). And the median recall under the three conditions was 0.14, 0.60, and 0.56, respectively ($\mu$=0.18, 0.61, 0.51, $Std$=0.19, 0.24, 0.28). An initial Kruskal-Wallis H-test indicates that precision and recall under the three conditions were significantly different (precision: $p$=0.002, effect size $\eta^2_H$=0.407, recall: $p$<0.001, $\eta^2_H$=0.511\footnote{The effect size of Kruskal-Wallis H-test $\eta^2_H$ was calculated according to \cite{tomczak2014need}.}). A post-hoc Dunn's test with Bonferroni correction ($\alpha$=0.05) showed that recall was improved significantly when comparing \textbf{C3} \vs \textbf{C1} and \textbf{C2} \vs \textbf{C1} (Figure \ref{fig:prec-recall}(c)). As for precision, \textbf{C2} was significantly higher than \textbf{C1}, while there was no sufficient proof to observe \textbf{C3} was higher than \textbf{C1}. We further analyzed the difference between \textbf{C2} and \textbf{C3}. On average, pathologists achieved $20.21\%$ higher recall and $21.51\%$ higher precision with \nap~than AI. However, there was no sufficient proof to observe that the precision and recall were significantly higher in  \textbf{C2} compared to \textbf{C3}.

It is noteworthy that participants' recall in identifying mitoses using the manual navigation is low. Upon further analysis of navigation traces, we found that the average mitotic rate in the areas participants visited with the manual navigation was 0.167/HPF (which is comparable to the average mitotic rate). As a comparison, the average mitotic rate with \nap~was 1.196/HPF, which is $6.17\times$ higher. We believe such a significant increase ($p$<0.001, $r$=0.851, Wilcoxon rank-sum test) in the prevalence rate of the target (\ie mitosis) is the main factor why \nap~could increase participants' recall: as described in \cite{wolfe2007low}, the low target prevalence would cause shifts of decision criteria that lead humans to miss targets in the visual search. \nap~harnesses AI to recommend highly-mitotic areas for users, which brings up the prevalence rate of the visual search targets, thus helping participants achieve higher recalls (even compared with AI).

High variances in precision and recall were observed when comparing \textbf{C2} and \textbf{C3}. We believe this was caused by two factors: \one variation in user interaction: in \textbf{C2}, participants chose a different recommendation customize settings and select a different amount of recommended ROIs in each trial (Figure \ref{fig:prec-recall}(f)-HPF). Variations in users interactions may also result in high variance in \textbf{C3} because the precision/recall in \textbf{C3} was calculated within the areas that participants visited in \textbf{C2}; \two Variation in user's experience: different participants might adapt different thresholds to call a cell as positive.

To conclude, \nap~achieved significantly higher precision and recall in identifying mitoses compared to manual navigation. Moreover, \nap, as a human + AI approach, might bring improvements compared to the AI-only condition: \nap~achieved higher precision and recall on average. However, we did not observe that such an improvement was statistically significant.

\begin{table*}
\caption{Summary of participants' questionnaire responses for the baseline and \nap~ with seven-scaled Likert questions. $p$ indicates the p-value of Wilcoxon test, and $r$ stands for the effect size. The numbers on the right indicate the averaged scores with their standard deviations. For Q1 -- Q5, 1=Not at all \dots 4=Neutral, \dots 7=Very. For Q6, 1=Very strongly prefer system 1 over system 2, 2=Strongly prefer system 1 over system 2, 3=Slightly prefer system 1 over system 2, \dots 4=Neutral, \dots, 7=Very strongly prefer system 2 over system 1.}
    \scalebox{1.0}{
    \begin{tabular}{clcccc}
    \toprule
        \textbf{ID} & \textbf{Question} & \textbf{Baseline} & \nap~ & $p$ & $r$\\
    \midrule
        Q1 & How hard did you have to work mentally to accomplish the tasks?& 5.13(1.30) & 2.93(1.10) & < 0.001 & 0.658\\
        Q2 & How would you describe your frustrations during the tasks? & 4.07(1.91) & 2.40(1.06) & 0.024 & 0.412\\
        Q3 & How capable is the system at helping count mitosis? & 2.79(1.63) & 6.43(0.65) & < 0.001 & 0.704 \\
        Q4 & How confident do you feel about your accuracy? & 4.21(1.42) & 5.93(0.73) & 0.004 & 0.530\\
        Q5 & Would you like to use the system in the future? & 4.13(1.92) & 6.47(0.64) & 0.001 & 0.594 \\
        Q6 & Overall Preference & \multicolumn{2}{c}{6.33(0.82)} & \multicolumn{2}{c}{N/A} \\
    \bottomrule
    \end{tabular}}
        \label{tab:question_response}
\end{table*}

\subsubsection{RQ2: Can \nap~ save pathologists’ time and effort?}

On average, participants spent 10min27s in each trial with the baseline system, and 13min8s with \nap. A Wilcoxon rank-sum test indicated no sufficient proof to conclude that participants' examinations were significantly longer ($p$=0.09, effect size $r$=0.306\footnote{The effect size of the Wilcoxon Test $r$ is calculated as $r=\frac{Z}{\sqrt{N}}$, where $Z$ is z-score from the Wilcoxon Test, and $N$ is the number of observations (30 in this study).}, Wilcoxon rank-sum test, same following). We further calculated each participant's navigation efficiency. The results showed that participants saw significantly more mitoses in unit time with \nap \\compared to manual navigation (manual: $\mu$=0.012 mitoses/second, \\  \nap: $\mu$=0.028 mitoses/second, $p$=0.002, $r$=0.579). Specifically, \nap's Local recommendations served as a shortcut that guided participants directly to highly-mitotic areas without manual searching:
\textit{``The local recommendations have more mitosis inside, and I can focus on this area. I can start counting from there and I do not need to find one myself.''}(P1)
\textit{``It (\nap) tells you which ones are the highest areas. And then you just go from there and decide. With system 1, you still have to review the whole slide.''}(P3)

In the post-study questionnaire, participants reported significantly less mental effort with \nap~(manual: $\mu=$5.13, \nap: $\mu=$2.93, $p$<0.001, $r$=0.658) compared to the manual navigation (Table \ref{tab:question_response} Q1). Furthermore, participants expressed less frustration using \nap~(manual: $\mu=$4.07,  \nap: $\mu=$2.40, $p$=0.024, $r$=0.412, Table \ref{tab:question_response} Q2). Specifically, participants valued \nap's Cell recommendations as the key to reducing the workload --- 
\textit{``It (\nap) takes away the burden of seeing and hunting for mitosis... it can tell you where is most likely to have mitosis and you decide `yes' or `no'.''}(P3)

In sum, although participants spent longer time using \nap~on average, their navigation efficiency was improved significantly by \nap's Local recommendations --- they could see more than twice the number of mitosis in unit time. Moreover, according to the questionnaire response, participants reported significantly less effort when using \nap. \nap's Cell recommendations contribute the main improvement: they could highlight specific cells from a large background, freeing pathologists from tedious manual visual search.

\subsubsection{RQ3: Compared to manual navigation, what is the benefit of using \nap?} 
\label{sec:rq3}
We answer this question by first comparing the patterns of interactions (\eg pan, zoom) while participants use \nap~ (\textbf{C2}) \vs with the manual navigation (\textbf{C1}). In sum, zooming and panning made up most of participants' interactions under \textbf{C1}, while ``selecting AI recommendations'' took the majority of interactions under \textbf{C2} (\nap). The median frequencies of zoom interactions under \textbf{C1} and \textbf{C2} were 37 and 6 (Figure \ref{fig:prec-recall}(d)). And the median pan interaction frequencies under \textbf{C1} and \textbf{C2} were 95 and 1 (Figure \ref{fig:prec-recall}(e)). A Wilcoxon test showed that zoom and pan interactions were significantly reduced under \textbf{C2} (zoom:$p$<0.001, $r$=0.651; pan: $p$<0.001, $r$=0.784). Furthermore, with \nap, participants selected a median of 6 Local, 27 HPF, and 8 Cell recommendations in each trial.

According to the questionnaire responses, participants believed that \nap~was more capable of assisting in detecting mitosis (manual: $\mu$=2.79, \nap:$\mu$=6.43, $p$<0.001, $r$=0.704, Table \ref{tab:question_response} Q3). Pathologists' confidence in mitosis reportings was improved significantly by \nap~(manual: $\mu$=4.21, \nap:$ \mu$=5.93, $p$=0.004, $r$=0.530, Table \ref{tab:question_response} Q4). Specifically, participants expressed that the AI recommendations would serve as a second opinion while they made justifications --- \textit{``I was kind of like 90\% sure ... but then if AI was 100\% sure, I felt more confident in saying that it was real mitoses.''}(P3). \textit{``It's kind of like having a second set of brains.''}(P6). Finally, participants expressed that they were more likely to use \nap~in the future (manual: $\mu$=4.13, \nap:$\mu$=6.47, $p$=0.001, $r$=0.594, Table \ref{tab:question_response} Q5). Overall, as shown in Table \ref{tab:question_response} Q6, participants indicated a preference for system 2 (\nap) over system 1 (baseline pathology scan viewer): based on the questionnaire, 8/15 of the participants rated a score 7 (very strongly prefer system 2 over system 1), 4/15 rated a score 6 (strongly prefer system 2 over system 1), and 3/15 rated a score 5 (slightly preferred system 2 over system 1).

In sum, users could navigate the pathology scans by selecting AI recommendations from \nap. Meanwhile, their pan and zoom interactions were significantly reduced. Overall, they believed \nap~was more capable of finding mitosis, had higher confidence while using \nap, and preferred to use it in the future.

\begin{figure*}
    \centering
    \includegraphics[width=1.0\linewidth]{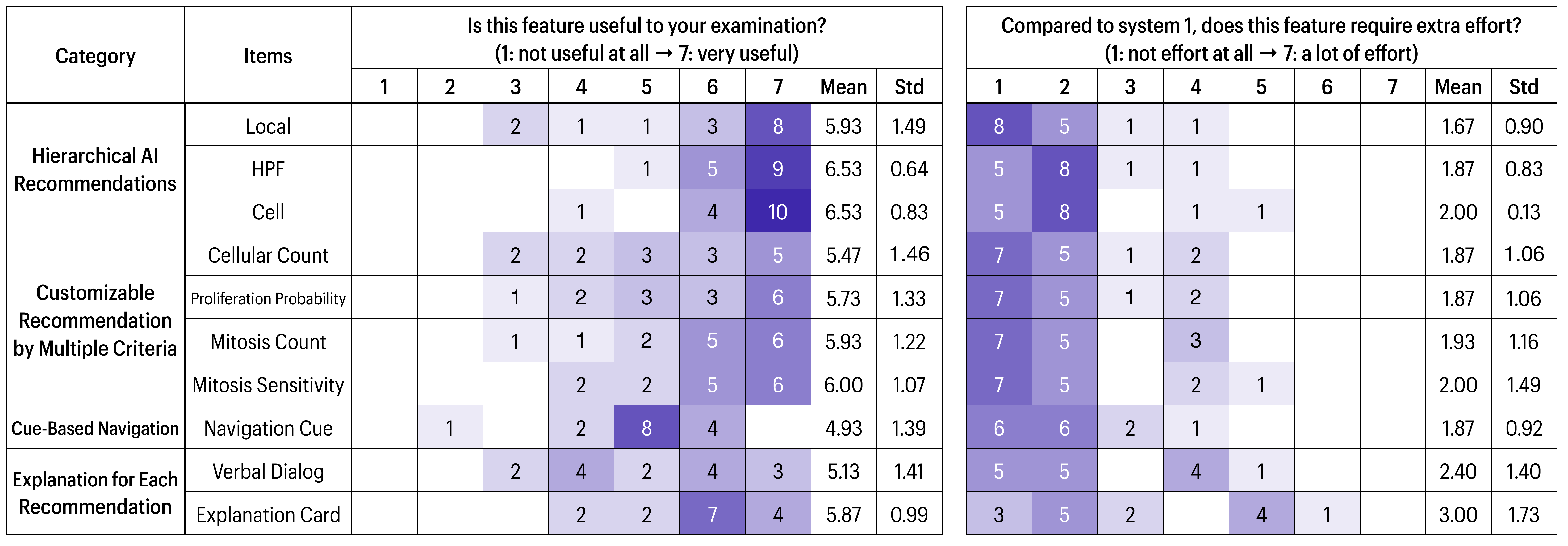}
    \caption{Participants' ratings on whether each component in \nap~is useful to pathologists' examination (left) / requires extra effort compared to the manual baseline system (system 1) (right).}
    \label{fig:util-rating}
    \Description{Two tables of participants' ratings on NaviPath's each component, including "Hierarchical AI recommendations", "Customizable Recommendation by Multiple Criteria", "Cue-Based Navigation", and "Explanation for Each Recommendation". The left table shows participants' overall ratings from the question: "Is this feature useful to your examination?" "1" stands for "not useful at all", and "7" means "very useful". The right table shows participants' overall ratings from the question: "Compared to system 1, does this feature require extra effort?" "1" stands for "no effort at all", and "7" means "a lot of effort".}
\end{figure*}

\subsection{Ratings on \nap's Components}

To further understand whether each \nap~component was useful for pathologists, we asked participants to rate each (see Figure \ref{fig:util-rating}). Here, we report the participants' ratings and discuss qualitative findings, organized by the categories of components:

\subsubsection{Hierarchical AI Recommendations} Participants rated average useful ratings of 5.93/7, 6.53/7, and 6.53/7 for Local, HPF, and Cell recommendations, respectively. Specifically, participants expressed that Local and HPF recommendations helped them narrow down from a large region without manual navigation --- \textit{``The entire slide might have thousands of high-power fields, and the Local recommendations picked the highest 36 for me ... the HPF recommendations continued to pick about 20 high-power fields from the Local recommendation ... it helps me rule out regions and focus on the important areas.''}(P14)

Notably, Cell recommendations received the highest useful rating among \nap's components. Participants expressed that Cell recommendations transformed the task of visual search into adjudication, which can save their mental effort. Specifically, they used Cell recommendations as an additional layer to quickly locate and adjudicate suspected cells: for most scenarios, participants directly reported the mitosis after glancing at the Cell recommendations. If they were not confident, they continued to select a Cell recommendation and examine it closely with a higher magnification. This explains why Cell recommendations were rated most useful, although they were not selected frequently in practice (as reported in Section \ref{sec:rq3}).
    
\subsubsection{Recommendation Customization by Multiple Criteria} Amongst the three criteria that \nap~used to generate recommendations, participants gave the ``mitosis count'' the highest usefulness rating ($\mu$=5.93/7), followed by the ``proliferation probability'' ($\mu$=5.73/7) and ``cellular count'' ($\mu$=5.47/7). Although most participants expressed that all three criteria should be considered in general, some (P2, P4, P15) believed it was not challenging for human pathologists to pick cellular areas, and it was not highly motivated to employ AI as such. 

We also found that participants did not frequently interact with the slide-bars to change the recommendation customization settings for the three criteria. Instead, they picked a custom set-up at the beginning of each trial and left them unchanged.  Upon further analysis, we found that \nap's recommendations might not change after users moved the slide-bars under certain circumstances, which disincentives users' interactions --- 
\textit{I don't see it (the recommendation) changing much when I set the `cellular count' as `high'.''}(P1) What's more, adjusting the customization settings during the examination might incur extra workload, and P14 suggested \nap~give pre-set values for the three criteria --- \textit{``It would be great if the system could give me default values for the three criteria ... changing the criteria is a lot of work if I see hundreds of slides.''}

Furthermore, participants had diverse opinions on how much a criterion should be considered in AI recommendations. One participant only gave ``mitosis count'' a high weight while giving zero weight for the other two criteria:
\textit{``I want AI to go straight to the mitoses, not like just predict for me based on the cell count where there are more mitoses elsewhere.''}(P4)
However, others thought \nap~should also include other criteria for recommendations. For example, P6 gave both ``cellular count'' and ``mitosis count'' a high weight ---
\textit{``I would like to include the cellular counts ... this is how we see tumors every day.''}(P6)

As for the sensitivity slide-bar, participants usually set it as ``high'' to see more recommendations, although this may produce false positives: \textit{``I move it all the way to the right, it will detect more mitosis  ... not all of them will be real mitosis, but it has more sensitivity. So then I can decide if the real to me or not.''}(P3) Pathologists' preferences of recall (sensitivity) over precision was also reported in a previous study \cite{gu2021xpath}. We believe such preferences are rooted from the imbalance risks in pathology decision making: while a proliferation of false-positive results (from low threshold) may cause longer time in examination, false-negative results (due to using a high threshold) might make the diagnosis unreliable because of the failure to acknowledge critical pathological features.

\subsubsection{Cue-Based Navigation} Surprisingly, the navigation cue received the lowest usefulness ratings by participants, with an average score of 4.93/7. Participants' opinions were split into two groups when asked how they used the navigation cue during work sessions. On one hand, some participants (P5, P10, P14) used cue-based navigation during their examination, and treated the navigation cue as a short-cut to access possible mitosis areas --- \textit{``It allows me to quickly locate the area where the next possible (mitosis) is located.''}(P5). On the other hand, some participants expressed that the cue-based navigation might be incompatible with a medical guideline:  \textit{``I sometimes did not know where these cues would guide me to ... because we need to see (mitoses in) 10 consecutive areas. And I didn't know if I was jumping from one to the other at the end they wouldn't be really consecutive'' }(P1) Regarding how participants might navigate under the high magnifications with \nap, we will discuss in more detail in Section \ref{sec:qual_hpf}.

\subsubsection{Explanations for Recommendations} Participants gave average ratings of 5.13/7 in usefulness and 2.40/7 in effort for the verbal explanation dialog. P5, P6, P7, P11, and P12 expressed that the verbal dialog assisted them in prioritizing the examination of HPF recommendations --- \textit{``Here (pointing at one HPF recommendation), it (the verbal dialog) says `very cellular' and `moderately likely'. And then here (pointing at another HPF recommendation), it says `very cellular' and `very likely'. So I might pick this box (the latter one) to see first ... it will be helpful to my selection.''}(P6) However, four participants (P10, P13, P14, P15) ignored the verbal dialog during the examination and used the ranking indexes to select HPF recommendations instead --- \textit{``I think the verbal dialog and the recommendation rankings are redundant ... the rule says the lower the (ranking) number, and more important the box is ... I feel that the ranking numbers are more straightforward.''}(P15)

As for the explanation card, participants gave a usefulness rating of 5.87/7. If participants were not confident about whether a Cell recommendation was mitosis, they would refer to the explanation card as a confirmation:
\textit{``I just took it as confirmatory that my assessment was correct.''} (P8) It is noteworthy that the explanation card also received the highest effort score (3.00/7) among \nap's components because participants spent extra effort comprehending the explanations.

\subsection{Qualitative Findings on Participants' Navigation Traces}

We analyzed participants' navigation traces on the pathology scans and report the qualitative findings on pathologists' navigation traces with the manual baseline system and \nap.

\subsubsection{Navigating the scan manually \vs with \nap} 
\begin{figure*}
    \centering
    \includegraphics[width=0.9\linewidth]{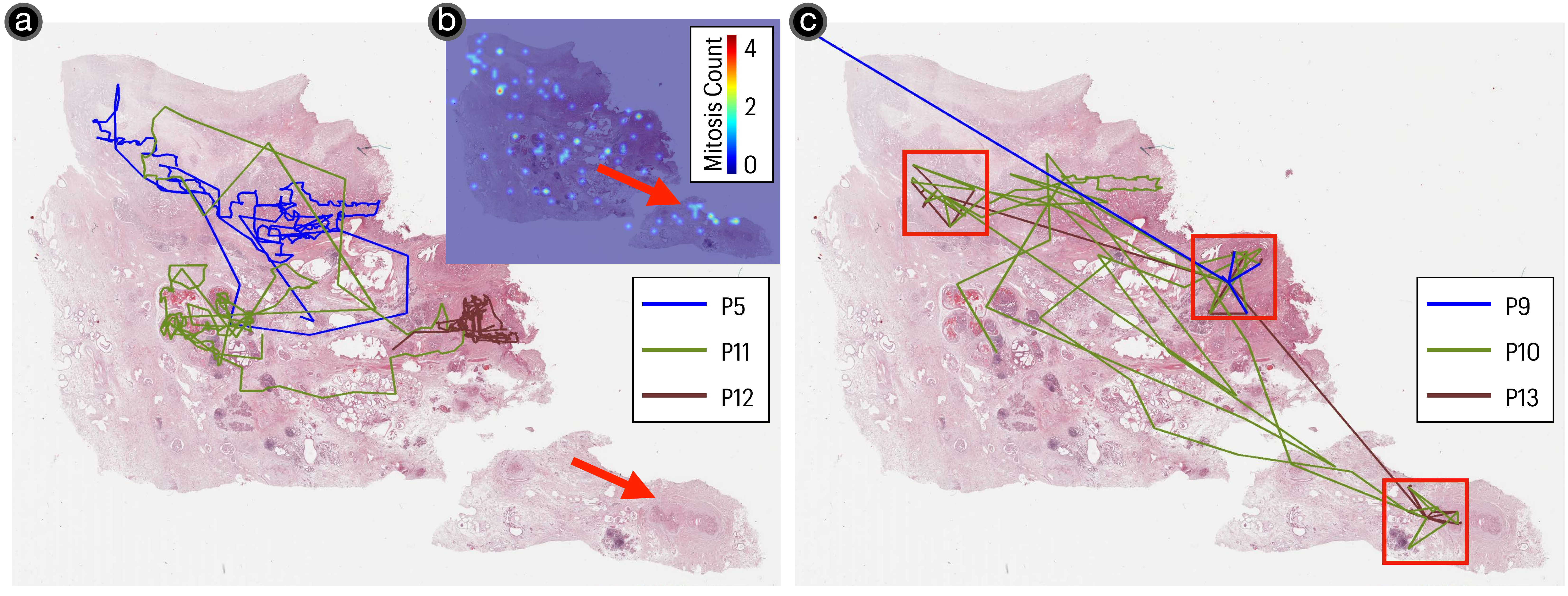}
    \caption{2D projections of participants' traces with manual and \nap~navigation on a pathology scan (zoom ignored). (a) Trace projections of P5, P11, and P12 with manual navigation. Note that all three participants did not examine the tissue on the bottom-right corner of the scan (pointed by the arrow). (b) The heatmap visualization of mitosis density of the scan. (c) Trace projections of P9, P10, and P13 with the \nap~navigation. The boxes highlight the approximate areas of Local recommendations generated by \nap.}
    \label{fig:navi_compr}
    \Description{Two subfigures showing the 2D projections of participants' navigation traces. The left figure shows three participants' (P5, P11, P12) navigation traces with the manual navigation. Their navigation traces barely overlap, and they do not cover an area in the corner that has aggregations of mitoses. The right figure shows the 2D projections of another three participants' (P9, P10, P13) navigation traces with NaviPath. Their navigation traces overlap more, thanks to NaviPath's AI recommendations. And two participants' (P10 and P13) navigation traces covers the corner area where P5, P11, P12 fail to examine with the manual navigation.}
\end{figure*}

One notorious issue of the pathology examination is the low between-subject consistency, which is usually caused by the randomness in pathologists' navigation. We also observed such randomness during our user study. For example, Figure \ref{fig:navi_compr}(a) visualizes the 2D projections of three (P5, P11, P12) participants' navigation traces with the manual navigation. It is noteworthy that all three traces barely overlap, which might result in inconsistencies in the medical decision makings. Also, all three participants did not examine a tissue session on the bottom-right corner of the scan (pointed by the arrow). However, according to the ground-truth mitosis density heatmap (Figure \ref{fig:navi_compr}(b)), the unexamined tissue session has aggregations of mitoses (shown as hotspots, pointed by the arrow). Therefore, the decisions made with the manual navigation might be biased because one important area was missed.

In contrast, participants' traces are more consistent with \nap. Figure \ref{fig:navi_compr}(c) illustrates three other participants' navigation traces (P9, P10, P13) within the same scan with \nap~navigation. The boxes indicate the approximate areas of Local recommendations generated by \nap. Thanks to AI recommendations, participants' navigation traces are more consistent within the three Local recommendations. Also, P10 and P13 examined the tissue session that had been missed in the manual navigation. 

\begin{figure*}
    \centering
    \includegraphics[width=0.9\linewidth]{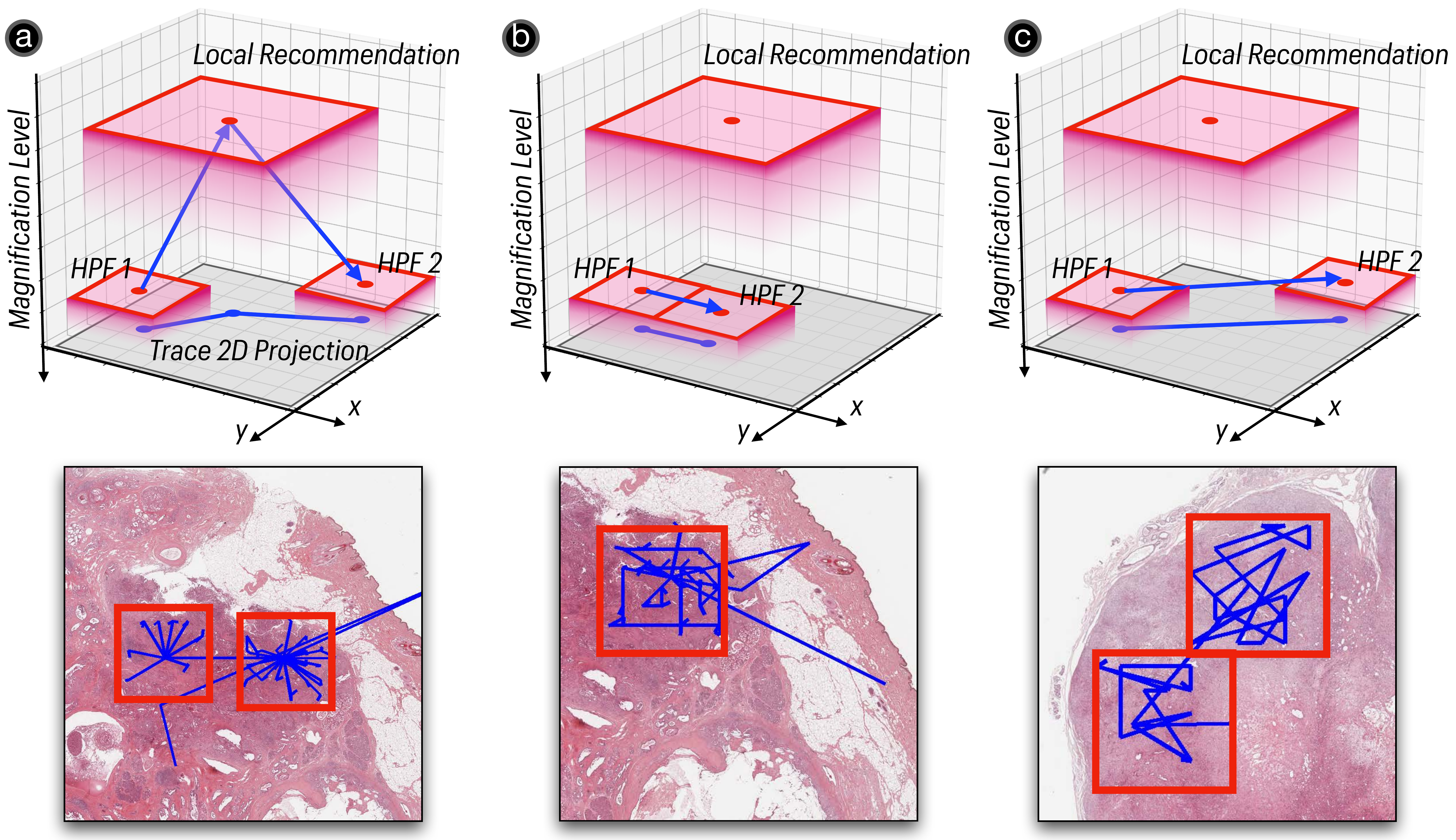}
    \caption{Three patterns of how our participants move to another HPF recommendation after examining one: (a) ``Diving'': first returned to the Local recommendation, overviewed the remaining HPF recommendations from the low magnification, and then dived down by selecting an HPF recommendation. The bottom figure shows 2D projections of participants' navigation traces during the work sessions; (b) ``Adjacent Panning'': directly pan to an adjacent HPF recommendation by clicking on the edge of \nap's interface; (c) ``Cue-Based Hopping'': directly hop to a remote HPF recommendation with the navigation cue.}
    \label{fig:navi_hpf}
    \Description{This figure shows 3D and 2D projections of participants' navigation traces of going to another HPF recommendation after finishing examining one with NaviPath. (a) The top figure shows a 3D illustration (z stands for the "magnification level") of a user's navigation trace for "diving": they first returned to the Local recommendation, overviewed the remaining HPF recommendations from the low magnification, and then dived down by selecting an HPF recommendation. The bottom figure shows 2D projections of participants' navigation traces during the work sessions, and a `spoke-like' navigation trace is resulted from the diving navigation. (b) "adjacent panning": a user directly pan to an adjacent HPF recommendation by clicking on the edge of NaviPath's interface. The adjacent panning results in a smooth, regular trace projection on the scan. (c) "cue-based hopping": a user directly hop to a remote HPF recommendation with the navigation cue. The navigation trace is more irregular under the cue-based hopping navigation.}
\end{figure*}

Therefore, \nap~can improve participants' consistency and also increase the exploration of their navigation. 

\subsubsection{Moving from one HPF recommendation to another with NaviPath.}
\label{sec:qual_hpf}
From the formative study, we learned that pathologists searched systematically in high magnifications with manual navigation. Here, we study whether our participants' navigation patterns in high magnifications with \nap~are different: specifically, we analyzed participants' navigation traces and summarized three navigation patterns of how our participants moved to another HPF recommendation after examining one:

\begin{itemize}
    \item \textbf{Diving}: Participants first moved to the Local recommendation, then overviewed remaining HPF recommendations with low magnification, and selected an HPF recommendation to examine in higher magnification (Figure \ref{fig:navi_hpf}(a)). During work sessions, P8 and P15 mainly used the diving navigation, and would switch the magnifications by selecting \nap's hierarchical recommendations without getting lost. As shown in Figure \ref{fig:navi_hpf}(a), the bottom figure, the diving navigation left a `spoke-like' navigation trace (the blue line) within each Local recommendation (red boxes).
    \item \textbf{Adjacent Panning}: Participants clicked on the edge of \nap's interface to move discretely to an adjacent HPF recommendation (Figure \ref{fig:navi_hpf}(b)). The adjacent panning is the closest to current pathologists' navigation practices (without AI), and five participants (P2, P3, P4, P7, P11) employed the adjacent panning in the study. The navigation trace is more regular with the adjacent panning (see Figure \ref{fig:navi_hpf}(b), the bottom figure).
    \item \textbf{Cue-Based Hopping}: Participants clicked on the navigation cue to hop to a remote HPF recommendation (Figure \ref{fig:navi_hpf}c). P5, P10, and P14 mainly used it during the study. With cue-based hopping, participants were able to see the HPF recommendations in ascending order based on ranking index to maximize navigation efficiency --- \textit{``My preference is to click on the navigation cue and jump to the next important HPF. For example, after I have seen number 1 (HPF recommendation), I will see number 2.''}(P10) As shown in Figure \ref{fig:navi_hpf}(c), the navigation trace is more irregular with cue-based hopping. 
\end{itemize}



\section{Discussion}
\subsection{Limitations}
\subsubsection{Limitations of the evaluation study}
\begin{itemize}
    \item \textbf{User Sampling}: The majority of participants are pathology residents with relatively less experience, making the conclusions for \textbf{RQ1} inevitably speculative due to a lack of participation of more-experienced attending pathologists;
    \item \textbf{Study Set-Up}: The work sessions were relatively brief because of the scarce availability of participants, and no clinical experiments were conducted because of strict regulations from US Food and Drug Administration (FDA);
    \item \textbf{Materials}: All pathology scans used in the study have the same tumor type because of the rare availability of public datasets. Therefore, they lack variability to reflect the real-world distribution of pathology data;
    \item \textbf{Choice of Baseline}: No comparison between \nap~and other human-AI systems was conducted because there is a lack of open-source systems for mitosis detection. There was also no comparison conducted with the optical microscope, pathologists' primary approach to see tumor specimens, due to the COVID-19 pandemic.
\end{itemize}

Therefore, future works should concentrate on conducting larger-scaled, longer-termed, in-the-wild studies to evaluate the influence of implementing a human-AI collaborative navigation system for pathologists.

\subsubsection{Limitations of NaviPath}

\begin{itemize}
    \item The two deep learning models for the proliferation probability and mitosis classification were trained from images of one tumor, and their performance on other tumors is unknown;
    \item The current cue-based navigation design used in \nap~(\ie citylight) cannot provide the distance information of off-screen recommendations, and might be incompatible with specific medical guidelines;
    \item The current recommendation customization algorithm was not predictable under certain circumstances;
    \item \nap~ does not support users to add their own ROIs for examination. Thus, users need to examine manually if an area was not recommended.
\end{itemize}

As such, future work should train AI models from various tumors to improve the model's generalizability. And future systems might consider other cue-based navigation designs (\eg Wedge \cite{gustafson2008wedge} or Halo \cite{10.1145/642611.642695}) that can offer both distance and directional information of off-screen targets, which can support navigation according to medical guidelines. Another improvement direction is modifying the overview map in the O+D design: by demonstrating where the pathologist is looking and all recommended ROIs to enhance humans' spatial awareness of off-screen targets (\eg \cite{bork2018towards}). Future works should also consider utilizing machine intelligence to support the examination of user-defined ROIs: for example, a user can select an area of interest manually, and the system can recommend all salient AI findings inside for the user to examine \cite{corvo2017pathova}. Finally, we also suggest future works to improve the predictability of medical AI, which we will discuss next.

\subsection{Implications for Human-AI Designs in Medical Decision-Making}

\subsubsection{Making AI-Enabled Systems Predictable} Previous work suggests that the disruptive behavior of AI might discourage medical professionals from using it in practice \cite{yang2016investigating}. In our study, we discovered that participants did not change the customization settings frequently because the outcomes were less predictable: for example, tuning the ``Cellular Count'' slide-bar would simultaneously change recommendations' locations and rankings. In some scenarios, tuning the slide-bar would not change the recommendations at all. 

It is challenging for doctors to be aware of whether the change is beneficial or the no change is caused by malfunction. As such, we suggest future human-AI systems in medicine to present intuitive clues that aid doctors in evaluating changes made by AI. For instance, future systems can justify why changes are happening or not -- text explanations generated by NLP agents (similar to \cite{wang2021brilliant}) can be implemented to explain the AI status and help pathologists comprehend the recommendation reasoning process. Another future direction might include making the recommendation AI less disruptive: for example, recommendations based on human-understandable medical concepts can make the algorithm more predictable for medical users \cite{cai2019human}. 

\subsubsection{Balancing Simplicity and Informativeness} Doctors prefer simple, straightforward designs \cite{10.1145/3449084}. From the evaluation study, we found that some participants preferred to use the ranking index number over the verbal explanation dialog. However, simpler designs usually mean ``lossy'' information compression, and might not be sufficiently informative for medical decision-making. Therefore, we suggest future HCI research to study what information should be preserved \vs discarded through empirical studies. For instance, Gu \etal~indicated that pathology AI systems could provide levels of AI explanations for doctors: a simple, visual explanation was shown by default, while more detailed explanations could be retrieved on demand \cite{gu2021xpath}. By balancing simplicity and informativeness, doctors can rapidly inquire about the most salient information with less confusion.

\subsubsection{Decoupling Doctors and AI} Recent research has reported that utilizing AI may cause doctors' diagnoses to align with that of AI's \cite{fogliato2022goes}. However, it is still unknown whether the alignment is beneficial or catastrophic because the performance of AI is subject to be influenced in clinical settings \cite{beede2020human}. Moreover, previous research suggests that the domain gap in pathology image data will harm AI performance \cite{stacke2020measuring, aubreville2021quantifying}. Therefore, doctors only examining within the AI-recommended areas would put physician-AI collaboration into a dilemma: on one hand, they may miss critical findings if the model's recall (sensitivity) is less than 1.00; on the other hand, seeing all areas comprehensively can barely reduce human workload. To tackle this problem of speed and accuracy, future improvements might consider re-designing the human-AI collaborative workflow: doctors might first overview a medical image and generate an overall impression of the case, then a human-AI collaborative system can be engaged to enable doctors to verify or refine their initial hypotheses \cite{10.1145/3449287}. What's more, providing additional sources of information might be an improvement: for example, attaching immunohistochemistry tests along with conventional pathology scans can let pathologists justify whether AI recommendations are reliable \cite{gu2021xpath}. Another unresolved question in this work is, since various pathological patterns might co-exist in a scan, are pathologists required to see other pathological patterns after examining one with \nap? In short, it depends on whether the criterion (in this work, mitosis) is deterministic for diagnoses according to the medical standard, and we suggest readers see \cite{gu2021xpath} for more detailed discussions.



\section{Conclusion}
This work introduces ~\nap~ to enhance pathologists' navigation efficiency in high-resolution tumor images by integrating domain knowledge and taking account of a practical workflow based on an empirical study with medical professionals. ~\nap~ could save pathologists from repetitive navigation in high-resolution tumor images through its AI-enabled designs. In contrast to prior work, we center on pathologists and adapt AI tools into their workflow to facilitate navigation processes. \nap~mainly focuses on mitosis in pathology, which represents a class of highly challenging problems on domain-specific navigation with high-resolution images. We hope insights provided by our solution can shed light on solving navigation challenges for other medical decision-making tasks.



\begin{acks}

This work was funded in part by the Young Investigator Award by the Office of Naval Research and the National Science Foundation under grant IIS-1850183. We appreciate the anonymous participants for the user study, and the reviewers for their valuable feedback in improving the manuscript.

\end{acks}


\balance
\bibliographystyle{ACM-Reference-Format}
\bibliography{references}

\end{document}